\documentclass[10pt]{article}

\usepackage{amsmath}
\usepackage{amssymb}

\usepackage{graphicx}

\usepackage{cite}

\usepackage{color} 

\usepackage[labelfont=bf,labelsep=period,justification=raggedright]{caption}

\makeatletter
\renewcommand{\@biblabel}[1]{\quad#1.}
\makeatother

\date{}

\begin{document}

\begin{flushleft}
{\Large
\textbf{Bayesian Test for Colocalisation Between Pairs of Genetic Association Studies Using Summary Statistics}
}
\\
Claudia Giambartolomei$^{1,\ast}$, 
Damjan Vukcevic$^{2}$,
Eric E. Schadt$^{3}$,
Lude Franke$^{4}$,
Aroon D. Hingorani$^{5}$, 
Chris Wallace$^{6}$,
Vincent Plagnol$^{1}$

\bf{1} UCL Genetics Institute, University College London (UCL), Darwin Building, Gower Street, London WC1E 6BT, UK
\\
\bf{2} Murdoch Childrens Research Institute, Royal Children's Hospital, Melbourne, Australia
\\
\bf{3} Department of Genetics and Genomics Sciences, Mount Sinai School of Medicine, New York, New York, United States of America
\\
\bf{4} Department of Genetics, University Medical Center Groningen, University of Groningen, Groningen, The Netherlands 
\\
\bf{5} Institute of Cardiovascular Science, University College London, London WC1E 6BT, UK
\\
\bf{6} JDRF/Wellcome Trust Diabetes and Inflammation Laboratory, Cambridge, Institute for Medical Research, Department of Medical Genetics, NIHR, Cambridge Biomedical Research Centre, University of Cambridge, Addenbrooke's Hospital, Hills Rd, Cambridge, CB2 0XY, UK
\\ 
$\ast$ E-mail: claudia.giambartolomei.10@ucl.ac.uk
\end{flushleft}

\section*{Abstract}
Genetic association studies, in particular the genome-wide association study (GWAS) design, have provided a wealth of novel insights into the aetiology of a wide range of human diseases and traits, in particular cardiovascular diseases and lipid biomarkers.
The next challenge consists of understanding the molecular basis of these associations.
The integration of multiple association datasets, including gene expression datasets, can contribute to this goal.
We have developed a novel statistical methodology to assess whether two association signals are consistent with a shared causal variant.
An application is the integration of disease scans with expression quantitative trait locus (eQTL) studies, but any pair of GWAS datasets can be integrated in this framework.
We demonstrate the value of the approach by re-analysing a gene expression dataset in 966 liver samples with a published meta-analysis of lipid traits including $> 100,000$ individuals of European ancestry \cite{Teslovich2010}.
Combining all lipid biomarkers, our re-analysis supported 29 out of 38 reported colocalisation results with eQTLs and identified 14 new colocalisation results, hence highlighting the value of a formal statistical test.
In two cases of reported eQTL-lipid pairs ({\it IFT172}, {\it TBKBP1}) for which our analysis suggests that the eQTL pattern is not consistent with the lipid association, we identify alternative colocalisation results with {\it GCKR} and {\it KPNB1}, indicating that these genes are more likely to be the causal in these genomic intervals.
A key feature of the method is the ability to derive the output statistics from single SNP summary statistics, hence making it possible to perform systematic meta-analysis type comparisons across multiple GWAS datasets (implemented online at  http://coloc.cs.ucl.ac.uk/coloc/).
Our methodology provides information about candidate causal genes in associated intervals and has direct implications for the understanding of complex diseases as well as the design of drugs to target disease pathways.

\section*{Author Summary}
Genome-wide association studies (GWAS) have found a large number of genetic regions (``loci'') affecting clinical end-points and phenotypes, many outside coding intervals. One approach to understanding the biological basis of these associations has been to explore whether GWAS signals from intermediate cellular phenotypes, in particular gene expression, are located in the same loci (``colocalise'') and are potentially mediating the disease signals. However it is not clear how to assess whether the same variants are responsible for the two GWAS signals or whether it is distinct causal variants close to each other. 
In this paper, we describe a statistical method which can use simply single variant summary statistics to test for colocalisation of GWAS signals. We describe one application of our method to a meta-analysis of blood lipids and liver expression, although any two datasets resulting from association studies can be used. 
Our method is able to detect the subset of GWAS signals explained by regulatory effects and identify candidate genes affected by the same GWAS variants.  As summary GWAS data are increasingly available, applications of colocalisation methods to integrate the findings will be essential for functional follow-up, and will also be particularly useful to identify tissue specific signals in eQTL datasets.

\section*{Introduction}

In the last decade, hundreds of genomic loci affecting complex diseases and disease relevant intermediate phenotypes have been found and robustly replicated using genome-wide association studies (GWAS, \cite{Feero2010}).
At the same time, gene expression measurements derived from microarray \cite{Nica2008} or RNA sequencing \cite{Pickrell2010} studies have been used extensively as an outcome trait for the GWAS design.
Such studies are usually referred to as  expression quantitative trait locus (eQTL) analysis.
While GWAS datasets have provided a steady flow of positive and replicable results, the interpretation of these findings, and in particular the identification of underlying molecular mechanisms, has proven to be challenging.
Integrating molecular level data and other disease relevant intermediate phenotypes with GWAS results is the natural step forward in order to understand the biological relevance of these results.

In this context, a natural question to ask is whether two independent association signals at the same locus, typically generated by two GWAS studies, are consistent with a shared causal variant.
If the answer is positive, we refer to this situation as colocalised traits, and the likelihood that both traits share a causal mechanism is greatly increased.
A typical example involves an eQTL study and a disease association result, which points to the causal gene and the tissue in which the effect is mediated \cite{Nica2010, Hunt2008, He2013}.
The same questions can also be considered between pairs of eQTLs \cite{Ding2010, Flutre2013}, or pairs of diseases \cite{Cotsapas2011}.

However, identifying the traits that share a common association signal is not a trivial statistical task.
Visual comparison of overlaps of association signals with an expression dataset is a step in this direction (using for example Sanger tool Genevar http://www.sanger.ac.uk/resources/software/genevar/), but the abundance of eQTLs in the human genome and across different tissues makes an accidental overlap between these signals very likely \cite{Nica2008}. 
Therefore visual comparison is not enough to make inferences about causality and formal statistical tests must be used to address this question.

Nica et al. \cite{Nica2010} proposed a methodology to rank the SNPs with an influence on two traits based on the residual association conditional on the most associated SNP. 
By comparing the GWAS SNP score with all other SNPs in the associated region, this method accounts for the local LD structure.
However this is not a formal test of a null hypothesis for, or against, colocalisation at the locus of interest. 
A formal test of colocalisation has been developed in a regression framework.
This is based on testing a null hypothesis of proportionality of regression coefficients for two traits across any set of SNPs, an assumption which should hold whenever they share causal variant(s) \cite{Plagnol2009, Wallace2012}.
No assumption is made about the number of causal variants, although the method does assume that in the case of multiple causal variants, all are shared.
However, both Nica's ranking method and proportionality testing share the drawback of having to specify a subset of SNPs to base the test on, and Wallace \cite{Wallace2013} shows that this step can generate significant biases. 
The main sources of bias are overestimation of effect sizes at selected SNPs (termed ``Winner's curse''), and the fact that, owing to random fluctuations, the causal variant may not always be the most strongly associated one.
These factors lead to rejection of colocalisation in situations where the causal SNP is in fact shared.
Although this can be overcome in the case of proportionality testing by averaging over the uncertainty associated with the best SNP models \cite{Wallace2013}, perhaps the greatest limitation is the requirement for individual level genotype data, which are rarely available for large scale eQTL datasets.

The success of GWAS meta-analyses has shown that there is considerable benefit in being able to derive association tests on the basis of summary statistics.
With these advantages in mind, He et al. \cite{He2013} developed a statistical test to match the pattern of gene expression with a GWAS dataset.
This approach, coded in the software Sherlock, can accommodate $p$-values as input.
However, their hypothesis of interest differs from the question of colocalisation, with the focus of the method being on genomewide convergence of signals, assuming an abundance of trans eQTLs.
In particular, SNPs that are not associated with gene expression do not contribute to the test statistic.
Such variants can provide strong evidence against colocalisation if they are strongly associated with the GWAS outcome. 

These limitations motivate the development of novel methodologies to test for colocalisation between pairs of traits.
Here, we derive a novel Bayesian statistical test for colocalisation that addresses many of the shortcomings of existing tools.
Our analysis focuses on a single genomic region at a time, with a major focus on interpreting the pattern of LD at that locus.

Our underlying model is closely related to the approach developed by Flutre et al. \cite{Flutre2013}, which considers the different but related problem of maximising the power to discover eQTLs in expression datasets of multiple tissues.
A key feature of our approach is that it only requires single SNP $p$-values and their minor allele frequencies (MAFs), or estimated allelic effect and standard error, combined with closed form analytical results that enable quick comparisons, even at the genome-wide scale.
Our Bayesian procedure provides intuitive posterior probabilities that can be easily interpreted. 
A main application of our method is the systematic comparison between a new GWAS dataset and a large catalogue of association studies in order to identify novel shared mechanisms.
We demonstrate the value of the method by re-analysing a large scale meta-analysis of blood lipids \cite{Teslovich2010} in combination with a gene expression study in 966 liver samples \cite{Schadt2012}.

\section*{Results}
\subsection*{Overview of the method}
We consider a situation where two traits (denoted by $Y_1, Y_2$) have been measured in two distinct datasets of unrelated individuals.
We assume that samples are drawn from the same ethnic group, i.e. allele frequencies and pattern of linkage disequilibrium (LD) are identical in both populations.
We consider, for each variant, a linear trend model between the outcome phenotype $Y$ and the genotypes $X$ (or a log-odds generalised linear model if one of the two outcome phenotypes $Y$ is binary):
$$Y = \mu + \beta X + \epsilon$$
We are interested in a situation where single variant association $p$-values and MAFs, or estimated regression coefficients $\hat{\beta}$ and their estimated precisions $var(\hat{\beta})$, are available for both datasets at $Q$ variants, typically SNPs but also indels.
We make two additional assumptions and discuss later in this paper how these can be relaxed. 
Firstly, that the causal variant is included in the set of $Q$ variants, either directly typed or well imputed \cite{Marchini2010, Howie2011, Howie2012}.
Secondly, that at most one association is present for each trait in the genomic region of interest.
We are interested in exploring whether the data support a shared causal variant for both traits.
While the method is fully applicable to case-control outcome, we consider two quantitative traits in this initial description.

\begin{figure}[!ht]    
  \begin{center}
    \includegraphics[width=6in]{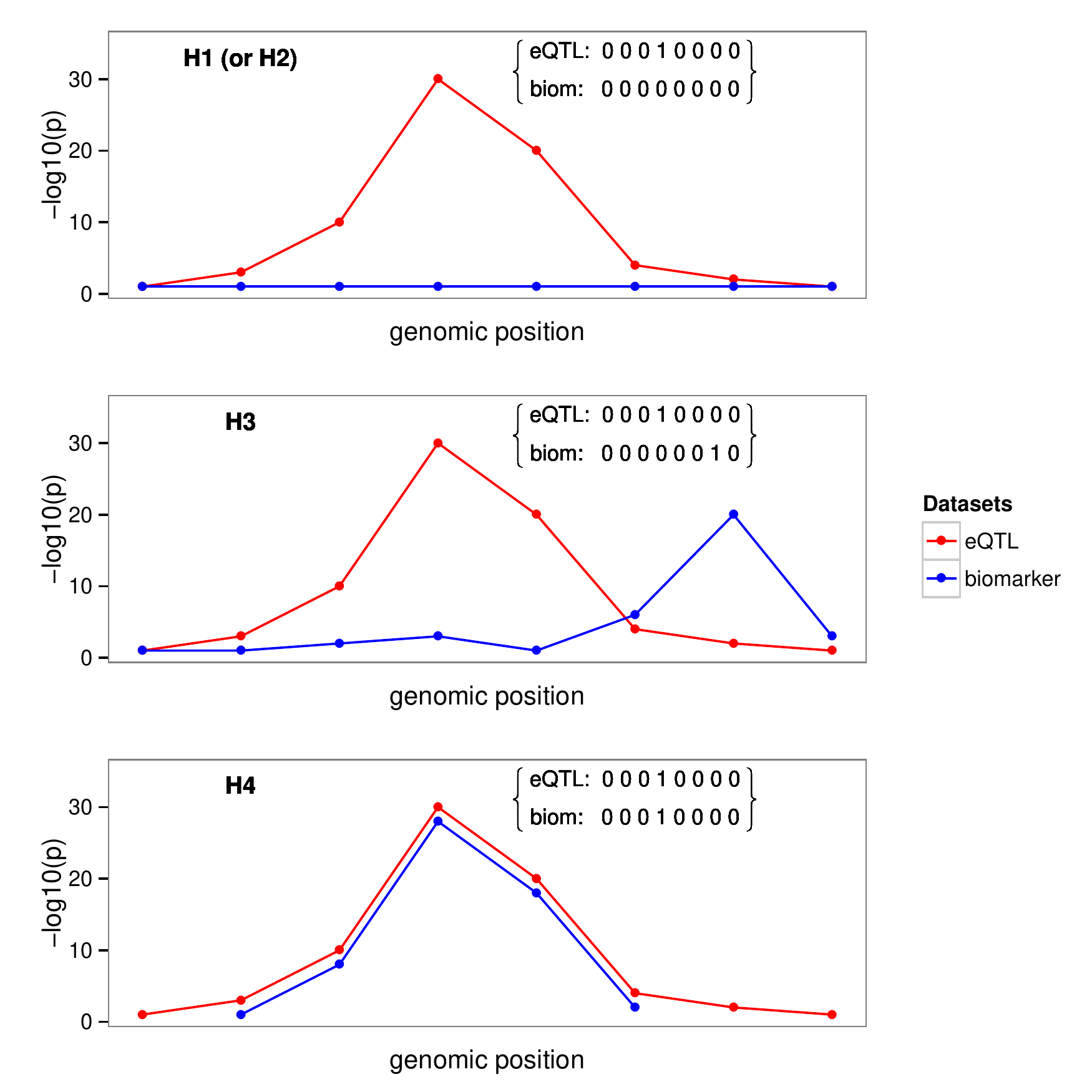}
  \end{center}
  \caption{
   {\bf Example of one configuration under different hypotheses represented by one binary vector for each trait of $(0,1)$ values of length $n = 8$, the number of shared variants in a region.} The value of $1$ means that the variant is causally involved in disease, $0$ that it is not. The first plot shows the case where only one dataset shows an association. The second plot shows that the causal SNP is different for the biomarker dataset compared to the expression dataset. The third plot shows the configuration where the single causal variant is the fourth one.
}
  \label{methods.figure}
\end{figure}

SNP causality in a region of Q variants can be summarised for each trait using a vector of length Q of (0, 1) values, where 1 means that the variant is causally associated with the trait of interest and at most one entry is non-zero.
A schematic illustration of this framework is provided in Figure \ref{methods.figure} in a region that contains 8 SNPs.
Each possible pair of vectors (for traits 1 and 2, which we refer to as ``configuration'') can be assigned to one of five hypothesis:
\begin{itemize}
\item $\mathbb{H}_0$: No association with either trait
\item $\mathbb{H}_1$: Association with trait 1, not with trait 2
\item $\mathbb{H}_2$: Association with trait 2, not with trait 1
\item $\mathbb{H}_3$: Association with trait 1 and trait 2, two independent SNPs
\item $\mathbb{H}_4$: Association with trait 1 and trait 2, one shared SNP
\end{itemize}
In this framework, the colocalisation problem can be re-formulated as assessing the support for all configurations (i.e. pairs of binary vectors) in hypothesis $\mathbb{H}_4$.

\begin{figure}[htb] 
\centering
   \includegraphics[trim={4cm 8.5cm 0cm 4.5cm},clip]{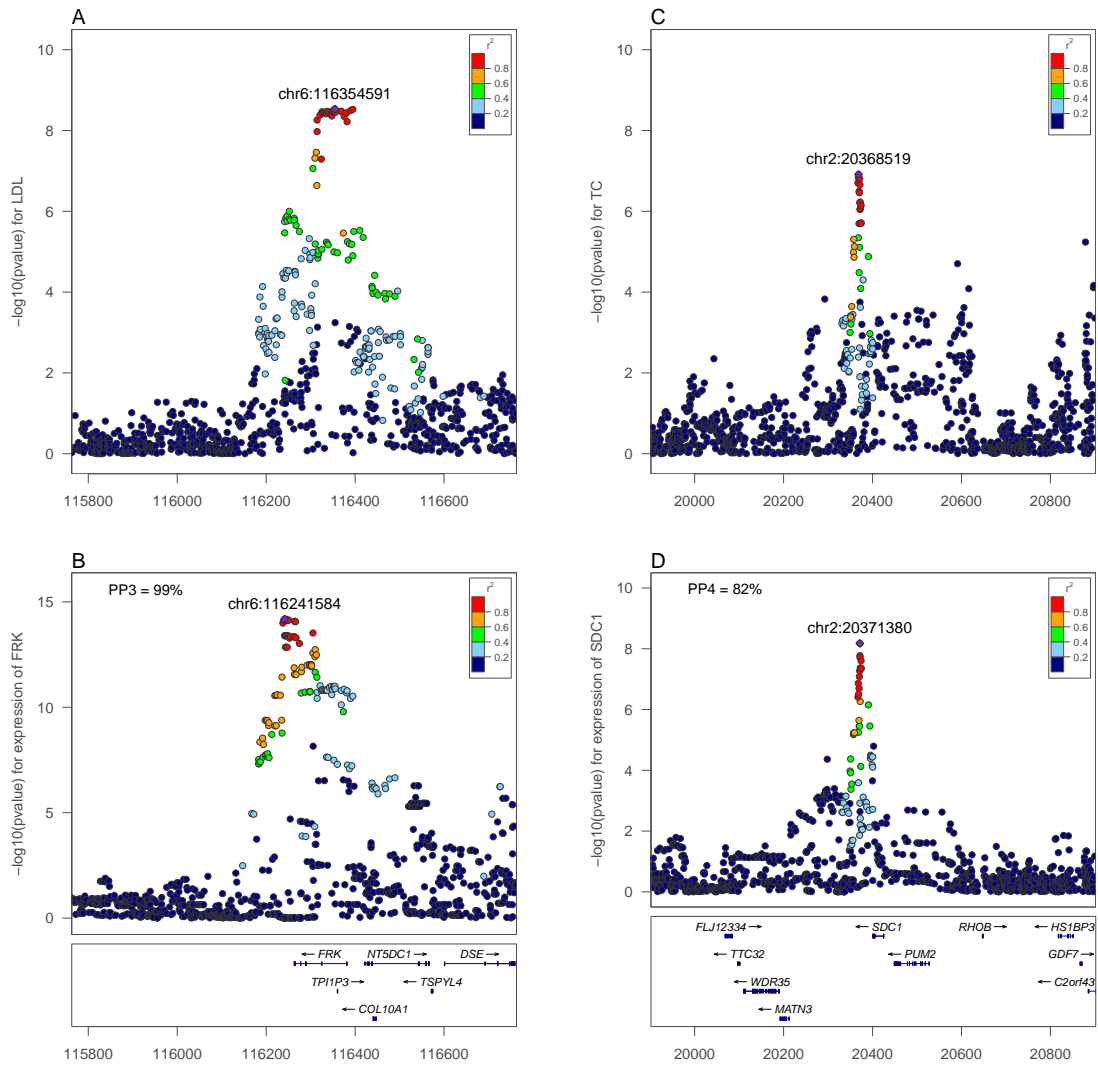}  \\
    \caption{
      \textbf{Illustration of the colocalisation results}. Negative(A-B, FRK gene and LDL, PP3 $>$ 90\%) and positive (C-D, SDC1 gene and total cholesterol, PP4 $>$ 80\%) colocalisation results. -log10(p) association $p$-values for biomarker (top, A and C) and -log10(p) association $p$-values for expression (bottom, B and D) at the {\it FRK} (A, B) and  {\it SDC1} locus (C, D), 1Mb range.
   }
  \label{fig_results}
\end{figure}

Our method is Bayesian in the sense that it integrates over all possible configurations.
This process requires the definition of prior probabilities which are defined at the SNP level (Methods).
A likelihood can be assigned to each configuration, and these likelihoods can be summed over all configurations and combined with the prior to assess the support for each hypotheses $(\mathbb{H})_1^5$
The result of this procedure is five posterior probabilities (PP0, PP1, PP2, PP3 and PP4).
A large posterior probability for hypothesis 3, PP3, indicates support for two independent causal SNPs associated with each trait.
In contrast, if PP4 is large, the data support a single variant affecting both traits.
An illustration of the method is shown in Figure \ref{fig_results} for negative (Fig \ref{fig_results}A-B, {\it FRK} gene and LDL, PP3 $>$ 90\%) and positive (Fig \ref{fig_results}C-D, {\it SDC1} gene and total cholesterol, PP4 $>$ 80\%) colocalisation results.

While the method uses Approximate Bayes Factor computations (ABF, \cite{Wakefield2009}, and Methods), no iterative computation scheme (such as Markov Chain Monte Carlo) is required.
Therefore, computations are quick and do not require any specific computing infrastructure.
Precisely, the computation time behaves as $Q^d$, where $Q$ is the number of variants in the genomic region and $d$ the number distinct associations (typically $d = 2$, assuming two traits and at most one causal variant per trait).
For all applications considered here, the computation time is nearly instant.
Importantly, the use of ABF enable the computation of posterior probabilities from single variant association $p$-values and MAFs, although the estimated single SNP regression coefficients $\hat{\beta}$ and their variances or standard errors are preferred for imputed data.

\bigskip

\subsection*{Sample size required for colocalisation analysis}
\begin{figure}[!ht]  
  \begin{center}
     \includegraphics[width=4in]{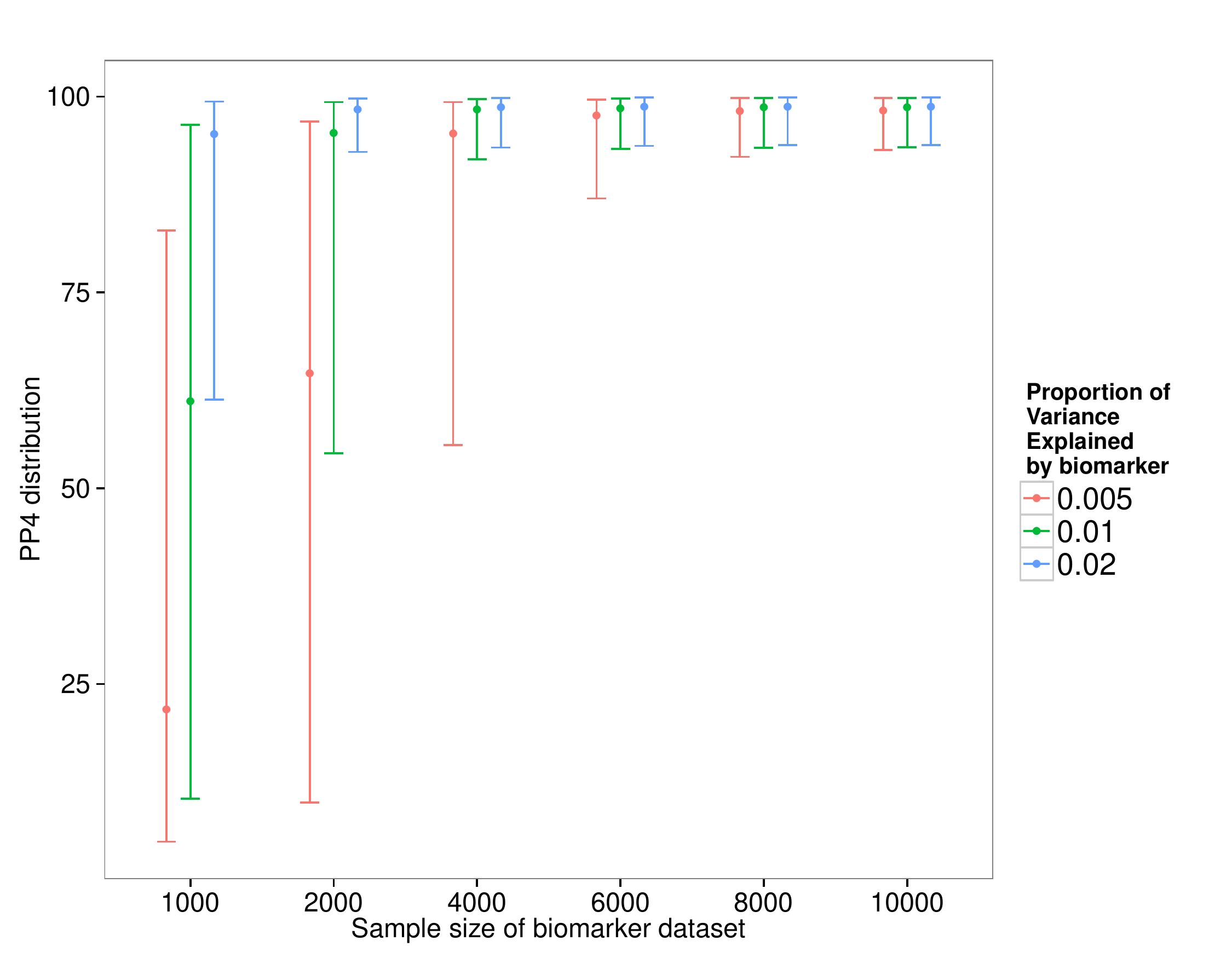}
  \end{center}  
  \caption{
   {\bf Simulation analysis with a shared causal variant between two studies: one eQTL (sample size 966 samples, 10\% of the variance explained by the variant) and one biomarker (such as LDL).}
    The variance explained by the biomarker is colour coded and the x-axis shows the sample size of the biomarker study.
    The y axis shows the median, 10\% and 90\% quantile of the distribution of PP4 values (which supports a shared common variant).
}
    \label{sample_size_simulations}
\end{figure}

Given the well understood requirements for large sample size for GWAS data, we used simulations to investigate the power of our approach.
We generated pairs of eQTL/biomarker datasets assuming a shared causal variant.
We varied two parameters: the sample size of the biomarker dataset and the proportion of the biomarker variance explained by the shared genetic variant.
We set the proportion of the eQTL variance explained by the shared variant to 10\% and we used the original sample size of the liver eQTL dataset described herein \cite{Schadt2012}.
Results are shown in Figure \ref{sample_size_simulations}.
We find that given a sample size of 2,000 individuals for the biomarker dataset, the causal variant needs to explain close to 2\% of the variance of the biomarker to provide reliable evidence in favour of a colocalised signal (lower $10^{th}$ percentile for PP4 $> 80\%$).

\subsection*{Consequence of limited variant density}
Until recently the assumption that, for a given GWAS signal, the causal variant in that interval had been genotyped was unrealistic.
However, the application of imputation techniques \cite{Marchini2010, Howie2011, Howie2012} can provide genotype information about the majority of common genetic variants.
Therefore, in situations where a common variant drives the GWAS signal, it is now plausible that, in imputed datasets, genotype information about this variant is available.
Nevertheless, limited imputation quality can invalidate this hypothesis.
This prompted us to investigate the implication of not including the causal variant in the genotype panel.

\begin{figure}[!ht]  
  \begin{center}
    \includegraphics[width=6in]{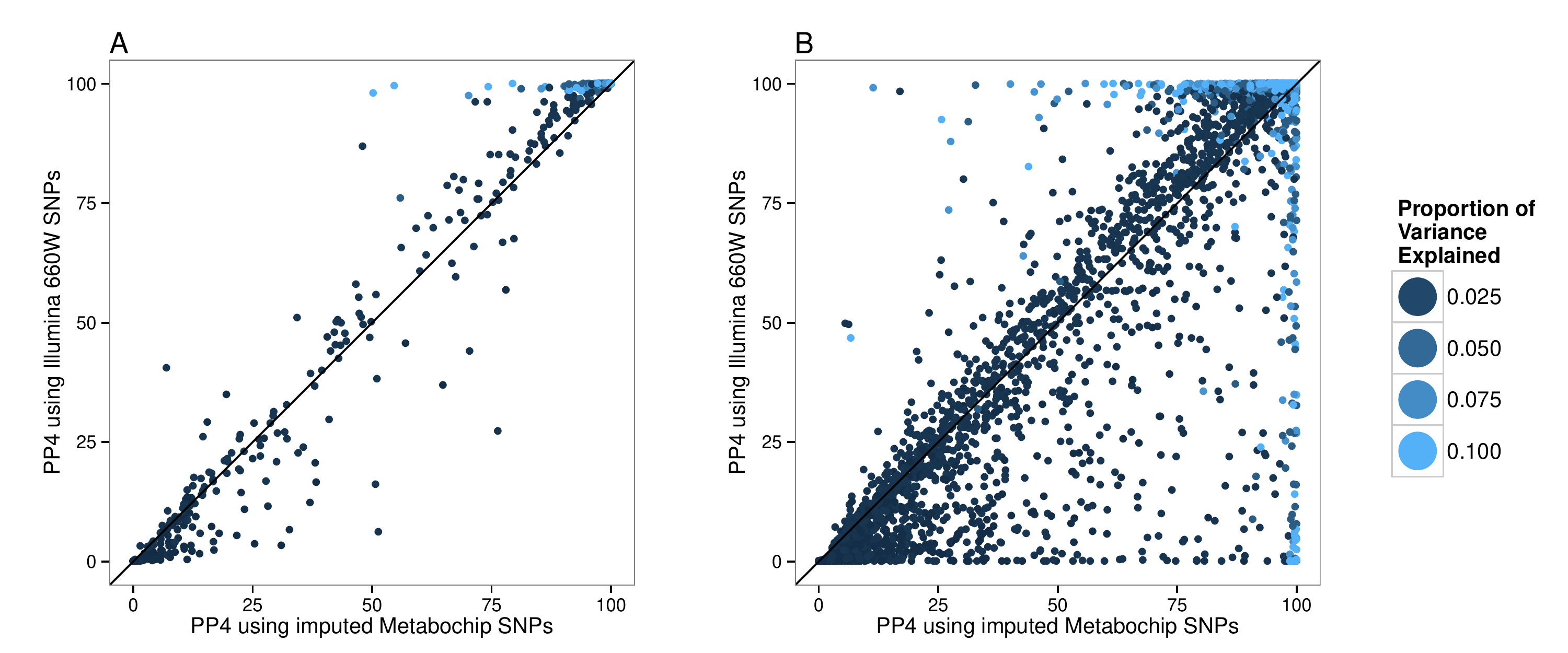}
  \end{center}
  \caption{
   {\bf Simulation analysis with a shared causal variant between two studies: one eQTL (sample size 966 samples) and one biomarker (sample size of 4,000 samples).} 
     The variance explained by the biomarker and the expression is the same and is colour coded. The x-axis shows the estimated PP4 for 1,000 simulations using data imputed from metaboChip Illumina array. 
    The y-axis uses the same dataset restricted to variants present on the Illumina 660W genotyping array to assess the impact of a lower variant density.
    \textbf{A}. The causal variant is included in the Illumina 660W panel. \textbf{B}. The causal SNP not included in Illumina 660W panel.
    }
    \label{compare_density}
\end{figure}

To address this question, we used Illumina MetaboChip data and imputed the genotyped regions using the Minimac software (\cite{Howie2012} and Methods).
We then selected only the subset of variants present in the Illumina 660K genotyping array.
We simulated data under the assumption of a shared causal variant, with 4,000 individuals in the biomarker dataset.
We then computed the PP4 statistic with and without restricting the SNP set to the Illumina 660K Chip SNPs (Figure \ref{compare_density}).
We also considered two different scenarios, with the causal SNP included/not included in the Illumina 660W panel (Figure \ref{illumina_vs_meta_causal} and Figure \ref{illumina_vs_meta_nocausal} for more exhaustive simulations).

\begin{figure}[!ht]
\includegraphics[width=6in]{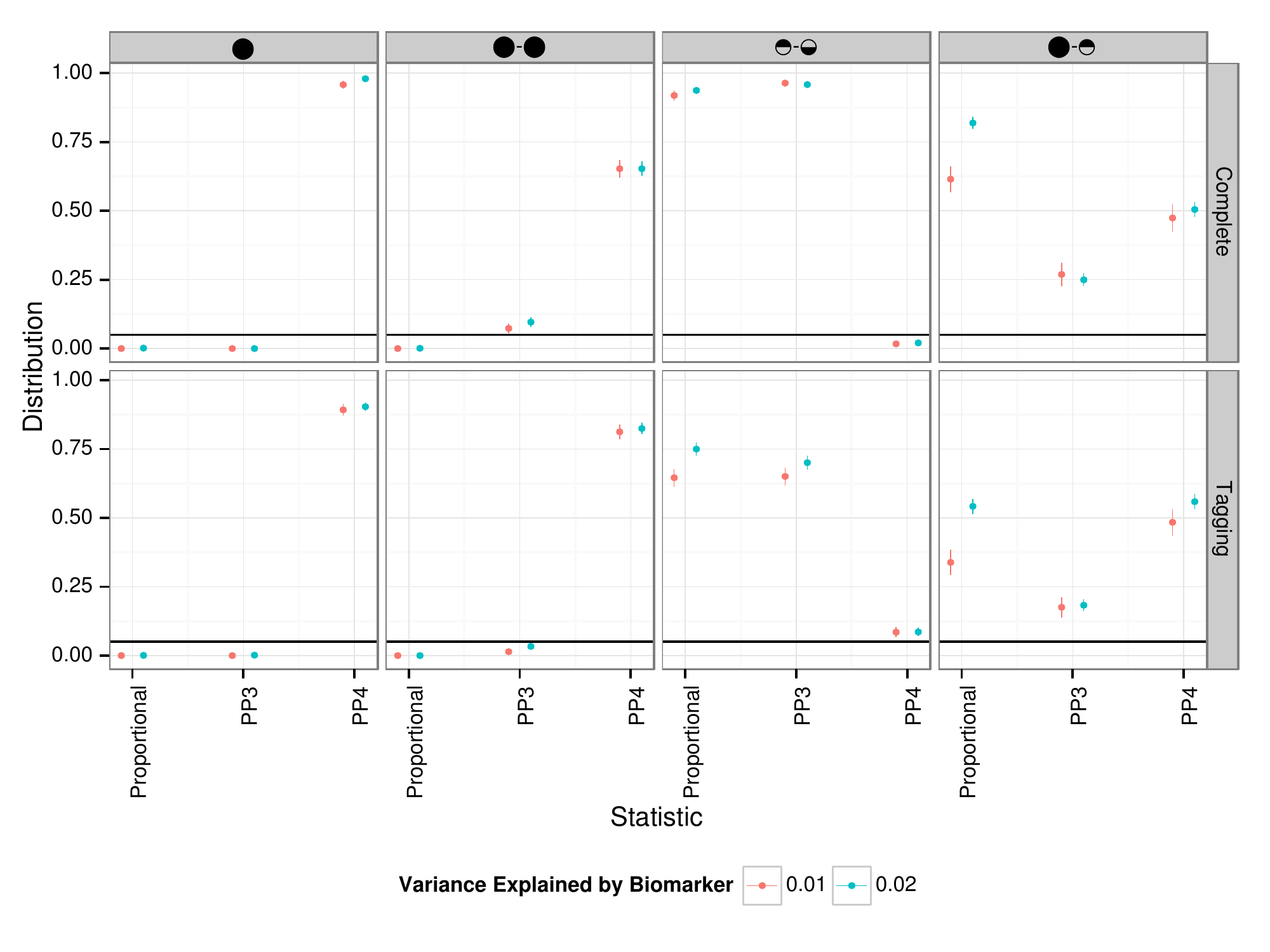}
  \caption{
   {\bf Summary of proportional and Bayesian colocalisation analysis of simulated data.} 
    Each plot shows a different scenario, the total number of causal variants in a region is indicated by number of circles in the plot titles with causal variants affecting both traits, the eQTL trait only, or the biomarker trait only, indicated by full circles, top-shaded circles and bottom-shaded circles respectively. In the top row the causal variant is typed or imputed, whereas only tag variants are typed/imputed in the bottom row. For proportional testing (under the BMA approach), we show the proportion of simulations with posterior predictive P-value $p < 0.05$ (black horizontal line) while for our Bayesian analysis we plot the proportion of simulations with the posterior probability (PP3 or PP4) of the indicated hypothesis $> 0.9$. Error bars show 95\% confidence intervals (estimated based on an average of 1,000 simulations per scenario). In all cases, for the eQTL sample size is 1,000; genetic variants explain a total of 10\% of eQTL variance; for the biomarker trait, the sample size is 10,000. }
  \label{compare_methods}
  \end{figure}

Our results show that when the causal variant is directly genotyped by the low density array, the use of imputed data is not essential (Figure \ref{compare_density}A).
However, in cases where the causal variant is not typed or imputed in the low density panel, the variance of PP4 is much higher (Figure \ref{compare_density}B).
In this situation, the resulting PP4 statistic tends to decrease even though considerable variability is observed.
Inspection of simulation results in Figure \ref{compare_methods} (bottom row for tagging SNP, leftmost graph for shared causal variant) shows that while PP4 tends to be lower than for its counterpart with complete genotype data (top row, leftmost graph), PP3 remains low.
This indicates that more probability is given to PP0, PP1 and PP2, which can be interpreted as a loss of power rather than misleading inference in favour of distinct variants for both traits.

\subsection*{Comparison with existing colocalisation tests}
We compared the behaviour of our proposed test with that of proportional colocalisation testing \cite{Plagnol2009, Wallace2013} in the specific case of a biomarker dataset with 10,000 samples (Figure \ref{compare_methods}).
Broadly, in the case of either a single common causal variant or two distinct causal variants, our proposed method could infer the simulated hypotheses correctly (PP4 or PP3 $>0.9$) with good confidence, and PP3 $>0.9$ slightly more often than the proportional testing $p$-value $<0.05$.
A key advantage in our Bayesian approach is the ability to distinguish evidence for colocalisation (i.e. high PP4) from a lack of power (i.e. high PP0, PP1 or PP2).
In both of these cases (high PP4 or high PP0/PP1/PP2), the use of the proportional approach leads to failure to reject the null even though the interpretation of these situations should differ.

It has been proposed that gene expression may be subject to both global regulatory variation which acts across multiple tissues and secondary tissue specific regulators \cite{Brown2012}.
Neither approach covers this case explicitly in its construction, but it is instructive to examine their expected behaviour. 
The proportional approach tends to reject a null of colocalisation, suggesting that a single distinct causal variant can be sufficient to violate the null hypothesis of proportional regression coefficients.  In contrast, the Bayesian approach tends to favour the shared variant in the cases covered by our simulations (median PP4 $>$ median PP3), and either hypotheses H3 or H4 can potentially have strong support (PP4 $>0.9$ in close to 50\% of simulations, and PP3 $>0.9$ in around 25\% of simulations). 
Of course, the ultimate goal should be to extend these tests to cover multiple causal variants, but in the meantime, it can be useful to know that a high PP4 in our proposed Bayesian analysis indicates strong support for ``at least one causal variant'' and that rejection of the null of proportionality of regression coefficients indicates that the two traits do not share all causal variants, not that they cannot share one.

\subsection*{Dealing with several independent associations for the same trait}  

We have so far assumed that each trait is associated with at most one causal variant per locus.
However, recent studies have shown that it is not unusual to observe two or more independent associations at a locus for a trait of interest \cite{Trynka2011}.
The natural and statistically exact modification of our approach would compute, for each trait, Bayes factors for sets of SNPs rather than single SNPs (up to $n$ SNPs jointly to accommodate for $n$ distinct associations per trait).
However, this approach has two drawbacks. 
Firstly, the interpretation of the resulting posterior probabilities is more challenging in situations where some but not all of the variants are shared across both traits.
More importantly, the typical approach consists of publishing single variant summary statistics, which would prevent the use of standard summary statistics, a key feature of our approach.

An alternative approach to account for multiple associations is motivated by the fact that, when distinct associations for the same trait are present at a locus, the assumption of a single variant per trait prompts the algorithm to consider only the strongest of these distinct association signals.
Secondary signals, including potential rare variant associations yielding marginal P-values, have little to no impact on the resulting posterior probabilities.
Therefore, a stepwise strategy successively conditioning on the top SNP to reveal the secondary association signals, can be an effective approximation.
In situations where only single SNP summary statistics are available, the approximate conditional meta-analysis framework proposed by Visscher et al \cite{Yang2012} can be used to obtain conditional $p$-values.
Another consequence of the behaviour of our inference procedure when the single causal variant assumption is enforced is that the posterior probabilities can be affected by the presence of two independent associations at the same locus with similar levels of significance.

\subsection*{Application to a meta-analysis of blood lipids combined with a liver expression dataset}
Teslovich et al. \cite{Teslovich2010} reported common variants associated with plasma concentrations of low-density lipoprotein cholesterol (LDL), high-density lipoprotein cholesterol (HDL) and triglyceride (TG) levels in more than 100,000 individuals of European ancestry. 
They then reported the correlations between the lead SNPs at the loci they found and the expression levels of transcripts in liver.
For the lipid dataset we have access only to summary statistics.
The liver expression dataset used in this analyses is the same as the one used in Teslovich et al. 
Teslovich et al. defined a region within 500 kilobases of the lead SNP, and the threshold for significance is $10^{-8}$. 
At this threshold, they found 38 SNP-to-gene eQTLs in liver (Supplementary Table 8 of \cite{Teslovich2010}). 
Table S1 shows our results for these 38 previously reported colocalisations. 
A complete list of all our identified colocalisations (independently of previous reports) is provided in Tables S2, S3, S4, S5 (broken down by lipid trait).

The majority of our results are consistent the findings of Teslovich et al (Table S1).
However, Table \ref{compare.tesl} lists the previously reported lipid-eQTL for which we find strong support {\emph against} the colocalisation hypothesis (PP3 $> 75\%$).
To assess the role of the prior, we varied the critical parameter $p_{12}$, which codes for the prior probability that a variant is associated with both traits.
Here we report the results using the $p_{12}=10^{-6}$. The complete list of results is provided in Table \ref{tesl_table}.
We found six potential ``false positive'' lipid-eQTL pairs for which our analysis does not support a colocalisation reported in Teslovich et al (Table \ref{compare.tesl}). 
In addition, we found strong evidence of distinct signals between  {\it HLA-DQ}/{\it HLA-DR} and TC (Table S1) but these results must be interpreted with caution owing to the extensive polymorphism in the  major histocompatibility complex region. 
Furthermore, the evidence for {\it TMEM50A} expression and LDL (PP3 = $58\%$) or TC (PP3 = $60\%$) is not strong enough to support either models (Table S1), so we cannot make a firm statement in this case.
In one case, {\it CEP250}, our re-analysis of the expression data did not identify an eQTL for this gene. 
In such a situation, both PP3 and PP4 are low and PP0, PP1 and PP2 concentrate most of the posterior distribution.

Table \ref{novel} lists the 14 colocalised loci (15 genes) that were not reported by Teslovich et al (or in Global Lipids Genetics Consortium \cite{Consortium2013} for the gene {\it NYNRIN}), but for which our method finds strong support for colocalisation (PP4 $> 75\%$).
Figure S6 shows the locuszoom plots for these colocalisation results.
Eleven of these 15 genes are strong candidates for involvement in lipid metabolism and/or have been previously suggested as candidate genes: {\it SDC1, TGOLN2, INHBB, UBXN2B, VLDLR, VIM, CYP26A1, OGFOD1, HP, HPR, PPARA}.
See supplemental for a brief overview of the function of these genes.
Four others genes have a less obvious link:  {\it CMTM6, C6orf106, CUX2, ENSG00000259359}. 

Three previously reported genes ({ \it SYPL2, IFT172, TBKBP1}) which, based on our re-analysis, do not colocalise with the lipid traits, have a nearby gene with a high probability of colocalisation (resp. {\it SORT1, GCKR, KPNB1}).
This suggests that these genes are more likely candidates in this region.
To explore the possibility that secondary signals may colocalise, we applied the stepwise regression strategy described above to deal with several independent associations at a single locus.
We performed colocalisation test using eQTL results conditional on the top eQTL associated variant.
Two of the loci ({\it SYPL2}/LDL or TC,  {\it APOC4} and TG) showed evidence of colocalisation with expression after conditional analysis (Table \ref{compare.tesl}).

\begin{figure}[htb] 
\centering
    \includegraphics[trim={4cm 2.2cm 4cm 4.4cm},clip]{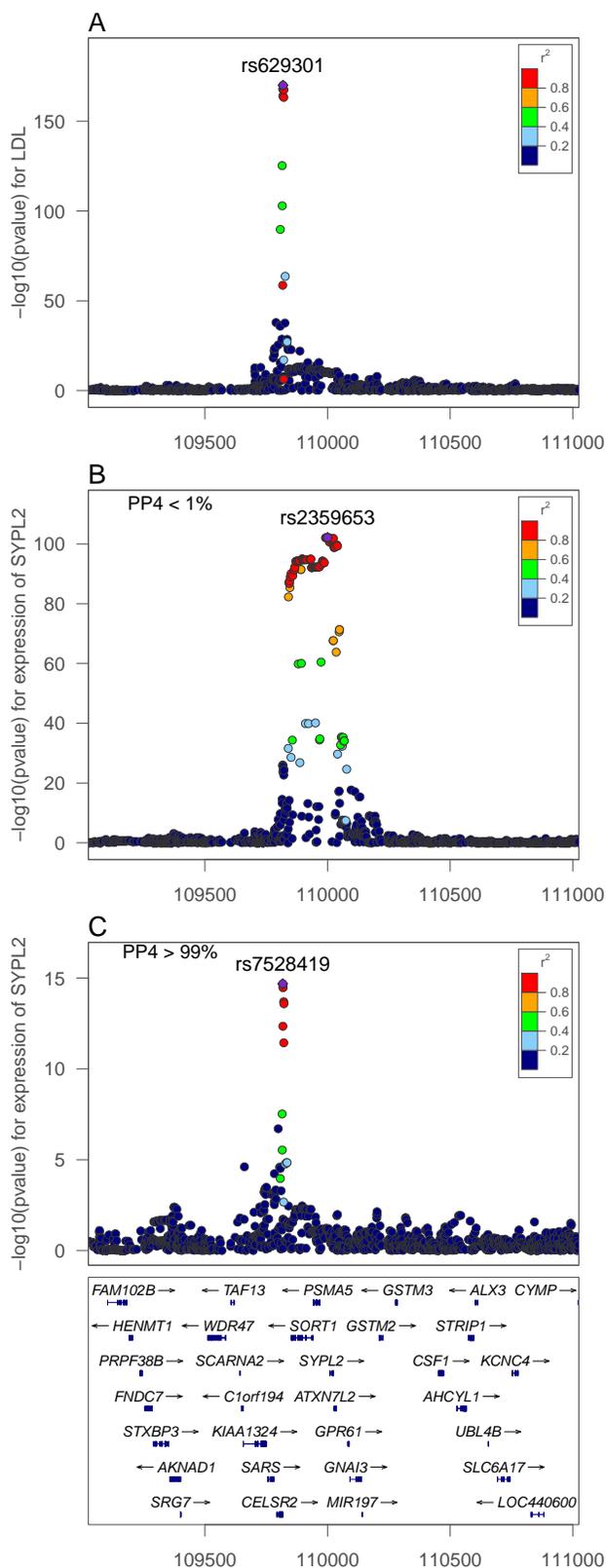} \\
  \caption{
   {\bf LDL association and eQTL association plots at the {\it SYPL2} locus.} 
    The x-axis shows the physical position on the chromosome (Mb)
    \textbf{A}: -log10(p) association $p$-values for LDL. The $p$-values are from the Teslovich et al published meta-analysis of $>$ 100,000 individuals.
    \textbf{B}: -log10(p) association $p$-values for {\it SYPL2} expression in 966 liver samples. 
    \textbf{C}: -log10(p) association $p$-values for {\it SYPL2} expression conditional on the top eQTL associated SNP at this locus (rs2359653).
   }
  \label{fig_sypl2}
\end{figure}

An example of this stepwise procedure for the gene {\it SYPL2} and LDL is provided in Figure \ref{fig_sypl2}.
We find that the top liver eQTL signal is clearly discordant with LDL association (Table \ref{compare.tesl} and Figure \ref{fig_sypl2}).
However, conditioning on the top eQTL signal reveals a second independent association for {\it SYPL2} expression in liver.
This secondary {\it SYPL2} eQTL colocalises with the LDL association (PP4 $>90\%$, Figure \ref{fig_sypl2}).

\subsection*{Web based resource}

We developed a web site designed for integration of GWAS results using only $p$-values and the sample size of the datasets (http://coloc.cs.ucl.ac.uk/coloc/). The website was developed using RWUI \cite{Newton2007}. 
Results include a list of potentially causal genes with the associated PP4 with their respective plots and ABF, and can be viewed either interactively or returned by email. 

Researchers can request a genome-wide scan of results from a genetic association analysis, and obtain a list of genes with a high probability of mediating the GWAS signals in a particular tissue. 
The tools also allows visualisation of the signals within a genetic region of interest.

The database and browser currently include the possibility of investigating colocalisation with liver \cite{Teslovich2010} and brain  \cite{Trabzuni2012, Ramasamy2013} expression data, however the resource will soon be extended to include expression in different tissues. 
This method, as well as alternative approaches for colocalisation testing \cite{Plagnol2009, Wallace2013}, are also available with additional input options in an R package, coloc, from the Comprehensive R Archive Network (http://cran.r\-project.org/web/packages/coloc).

\section*{Discussion}
We have developed a novel Bayesian statistical procedure to assess whether two association signals are colocalised.
The strength of this approach lies in its speed and analytical forms, combined with the fact that it can use single variant $p$-values when only these are available.
Our method differs from a typical fine-mapping exercise in the sense that we are not interested in knowing which variant is likely to be causal but only whether a shared causal variant is plausible.

Our results show that to provide an accurate answer to the colocalisation problem, high density genotyping and/or accurate use of imputation techniques are key.
The quality of the imputation is another important parameter. 
Indeed, while the variance of the regression coefficient can be estimated solely on the basis of the minor allele frequency for typed SNPs and sample size (and the case control ratio in the case of a binary outcome) \cite{Guan2008, Marchini2010}, this ignores the uncertainty due to imputation.
Filtering out poorly imputed SNPs partially addresses this problem, with the drawback that it may exclude the causal variant(s).
Hence, providing estimates of the variance of the MLE, together with the effect estimates, will result in greater accuracy.
This additional option is available on the coloc package in R (http://cran.r\-project.org/web/packages/coloc). 

We currently assume that each genetic variant is equally likely a priori to affect gene expression or trait.
A straightforward addition to our methodology would consider location specific priors for each variant, which would depend for example on the distance to the gene of interest, or the presence of functional elements in this chromosome region \cite{Khatun2012, Gerstein2012}. 
Our computation of the BF also assumes that, under $\mathbb{H}_4$, the effect sizes of the shared variant on both traits are independent.
This could be modified if, for example, one compares eQTLs across different tissue types, or the same trait in two different studies.
\cite{Wen2011} has proposed a framework to deal with correlated effect sizes, and these ideas could potentially be incorporated in our colocalisation test.

Another prior related issue is the choice of prior probabilities for the various configurations.
For the eQTL analysis, we used a $10^{-4}$ prior probability for a cis-eQTL.
A more stringent threshold may be better suited for trans-eQTLs where the variants are further away from the gene under genetic control.
We also used a prior probability of $10^{-4}$ for the lipid associations. 
Although our knowledge about this is still lacking, this estimate has been suggested in the literature in the context of GWAS \cite{Stephens2009, Wakefield2009, Burton2007}. 
We assigned a prior probability of $1 \times 10^{-6}$ for $p_{12}$, which encodes the probability that a variant affects both traits.
It has been shown that SNPs associated with complex traits are more likely to be eQTLs compared to other SNPs chosen at random from GWAS platforms \cite{Nicolae2010}, and a higher weighting for these SNPs has been proposed when performing Bayesian association analyses \cite{Knight2011, Johansson2012}. Also, eQTLs have been shown to be enriched for disease-associated SNPs when a disease-relevant tissue is used \cite{Ding2010, Richards2011}. 
Our sensitivity analysis for the  $p_{12}$ parameter showed broadly consistent results (Table S1).
In cases where GWAS data are available for both traits, \cite{Flutre2013} show that it is possible to estimate these parameters from the data using a hierarchical model.
This addition is a possible extension of our approach.

The interpretation of the posterior probabilities requires caution.
For example, a low PP4 may not indicate evidence against colocalisation in situations where PP3 is also low.
It may simply be the result of limited power, which is evidenced by high values of PP0, PP1 and/or PP2. 
Morever, a high PP4 is a measure of correlation, not causality.
To illustrate this, one can consider the relatively common situation where a single variant appears to affect the expression of several genes in a chromosome region (as observed, for example, in the region surrounding the {\it SORT1} gene).
Several eQTLs will be colocalised, both between them and with the biomarker of interest.
In this situation one would typically expect that a single gene is causally involved in the biomarker pathway but the colocalisation test with the biomarker will generate high PP4 values for all genes in the interval.

We show that we can use conditional $p$-values to deal with multiple independent associations with the same trait at one locus.
While we found this solution generally effective, Wallace \cite{Wallace2013} points out that this top SNP selection for the conditional analysis can create biases, although the bias is small in the case of large samples and/or strong effects.
For difficult loci with multiple associations for both traits and available genotype data, it may be more appropriate to estimate Bayes factors for sets rather than single variants in order to obtain an exact answer.
This extension would avoid the issue of SNP selection for the conditional analysis. 

Importantly, GWAS signals can be explained by eQTLs only when the causal variant affects the phenotype by altering the amount of mRNA produced, but not when the phenotype is affected by changing the type of protein produced, although the former seems to be the most common \cite{Nicolae2010}. 
Furthermore, since many diseases manifest their phenotype in certain tissues exclusively \cite{Nica2008,  Brown2012, Dimas2009, Hernandez2012}, colocalisation results will be dependent on the expression dataset used.
In addition to identifying the causal genes, the identification of tissue specificity for the molecular effects underlying GWAS signals is a key outcome of our method.
We anticipate that building a reference set of eQTL studies in multiple tissues will provide a useful check for every new GWAS dataset, pointing directly to potential candidate genes/tissue types where these effects are mediated.

While this report focuses on finding shared signals between a biomarker dataset and a liver expression dataset, we plan to utilise summary results of multiple GWAS and eQTL studies, for a variety of cell types and traits. 
In fact, our method can utilise summary results from any association studies. 
Disease/disease, ({\it cis} or {\it trans}) eQTL/disease or disease/biomarkers comparisons are all of biological interest and use the same statistical framework.
We expect that the fact that the test can be based on single SNP summary statistics will be key to overcome data sharing concerns, hence enabling a large scale implementation of this tool.
The increasing availability of RNA-Seq eQTL studies will further increase the opportunity to detect isoform specific eQTLs and their relevance to disease studies.
Owing to the increasing availability of GWAS datasets, the systematic application of this approach will potentially provide clues into the molecular mechanisms underlying GWAS signals and the aetiology of the disorders.

\section*{Materials and Methods}

\subsection*{Expression dataset}
We used in our analysis gene expression and genotype data from 966 human liver samples.
The samples were collected post-mortem or during surgical resection from unrelated European-American subjects from two different non-overlapping studies, which have been described in \cite{Schadt2012}.
The cohorts were both genotyped using Illumina 650Y BeadChip array, and 39,000 expression probes were profiled using Agilent human gene expression arrays. All of the expression data has been normalised as one unit even though they were part of different studies, since high concordance between data generated using the same array platforms has been previously reported. 
Probe sequences were searched against the human reference genome GRCh37 from 1000genomes using BLASTN. Multiple probes mapping to one gene were kept in order to examine possible splicing. 
The probes were kept and annotated to a specific gene if they were entirely included in genes defined by Ensembl ID or by HGNC symbol using the package biomaRt in R \cite{R}.  
After mapping and annotating the probes, we were left with 40,548 mapped probes covering 24,927 genes.

\subsection*{Imputation of genetic data}
Quality control filters were applied both before and after imputation. Before imputation, individuals with more than 10\% missing genotypes were removed, and SNPs showing a missing rate greater than 10\%, a deviation for HWE at a $p$-value less than 0.001 were dropped. After imputation, monomorphic SNPs were excluded from analyses. 
\\To speed up the imputation process, the genome was broken into small chunks that were phased and imputed separately and then re-assembled. This was achieved using the ChunkChromosome tool \linebreak (http://genome.sph.umich.edu/wiki/ChunkChromosome), and specifying chunks of 1000 SNPs, with an overlap window of 200 SNPs on each side, which improves accuracy near the edges during the phasing step.
Each chunk was phased using the program MACH1 (Versions 1.0.18, downloaded from: http://www.sph.umich.edu/csg/abecasis/MACH/download/) with the number of states set to 300 and the number of rounds of MCMC set to 20 for all chunks. Phased haplotypes were used as a basis for imputation of untyped SNPs using the program Minimac (version 2011.10.27, downloaded from: http://www.sph.umich.edu/csg/cfuchsb/minimac-beta-2012.8.15.tgz) with 1000 Genomes EUR reference haplotypes (phase1 version 3, March 2012) to impute SNPs not genotyped on the Illumina array. 
The data was then collated in probability format that can be used by the R Package snpStats\cite{R}.

\subsection*{eQTL analysis}

eQTL $p$-values, effect sizes, and standard errors were obtained by fitting a linear trend test regression between the expression of each gene and all variants 200 kilobases upstream and downstream from each probe.
After filtering out the variants with MAF $<$ 0.001, monomorphic SNPs, multi-allelic SNPs (as reported in 1000 Genomes or in the Ensembl database) and variants not sufficiently well imputed (Rsq $< 0.3$, as defined by minimac http://genome.sph.umich.edu/wiki/minimac) between both datasets, we applied our colocalisation procedure.
We conducted conditional analysis on SNPs with $p$-values $< 10^-4$ for the expression associations, and repeated the colocalisation test using expression data conditioned on the most significant SNP. The aim of this analysis is to explore whether additional signals for expression other than the main one are shared with the biomarker signal.

\subsection*{Biomarker dataset}
The biomarker $p$-values from the meta-analyses (with genomic control correction) were obtained from a publicly available repository (http://www.sph.umich.edu/csg/abecasis/public/lipids2010/).

\subsection*{Posterior Computation}

\noindent We call a ``configuration" one possible combination of pairs of binary vectors indicating whether the variant is associated with the selected trait. We can group the configurations into five sets, $S_0, S_1, S_2, S_3, S_4$, containing assignments of all SNPs Q to the functional role corresponding to the five hypothesis $\mathbb{H}_0, \mathbb{H}_1, \mathbb{H}_2, \mathbb{H}_3, \mathbb{H}_4$. 

\noindent We can compute the posterior probabilities for each of these 5 hypothesis using a Bayesian approach. Firstly, we calculate the likelihood of each hypothesis,

\begin{equation}
L(H_h) = L(H_h \mid D) = \sum_{S \in S_h} P(D \mid S) P(S)
\label{overall.lik}
\end{equation}

\noindent where P(S) is the prior probability of a configuration, $P(D \mid S)$ is the probability of the observed data D under a given configuration (which we will refer to as the likelihood of this configuration), and the sum is over all configurations S which are consistent with a given hypothesis $H_h$, h=(1,2,3,4).  Thus, each configuration likelihood is weighted by the prior probability of that configuration.

\noindent Next, we compute the posterior probability by taking the ratio of the hypothesis likelihoods under the different hypotheses. For example, the posterior probability under hypothesis 4 is:

\begin{equation}
PP4 = \frac{ L(H_4) } { L(H_0)  + L(H_1) +  L(H_2) +  L(H_3) +  L(H_4) }
\label{ratio.of.lik1}
\end{equation}

\noindent If we also divide each probability by the baseline $L(H_0)$ we get:

\begin{equation}
PP4 = \frac{ \frac{L(H_4)}{L(H_0)} } { 1  + \frac{L(H_1)}{L(H_0)}  +  \frac{L(H_2)}{L(H_0)} +  \frac{L(H_3)}{L(H_0)} +  \frac{L(H_4)}{L(H_0)} }
\label{ratio.of.lik2}
\end{equation}

\bigskip

\noindent The ratios in the numerator and denominator of equation \ref{ratio.of.lik2} are:
\begin{equation}
\frac{L(H_h)}{L(H_0)} = \sum_{S\in{S_h}} \frac{P(D \mid S)}{P(D \mid S_0)} \times \frac{P(S)}{P(S_0)} 
\label{likelihood.incr} 
\end{equation}
\noindent The first ratio in this equation is a Bayes Factor (BF) for each configuration, and the second ratio is the prior odds of a configuration compared with the baseline configuration $S_0$. The BF can be computed for each variant from the $p$-value, or estimated regression coefficient $\hat{\beta}$ and variance of $\hat{\beta}$, using Wakefield's method. By summing over all configurations in $S_h$ we are effectively comparing the support in the data for one alternative hypothesis versus the null hypothesis.

\subsection*{Bayes factor computation}

A Bayes Factor for each SNP and each trait 1 and 2 was computed using the Approximate Bayes Factor (ABF, \cite{Wakefield2009}). Wakefield's method yields a Bayes factor that measures relative support for a model in which the SNP is associated with the trait compared to the null model of no association. 

\noindent The equation used is the following:
\begin{equation}
ABF = \frac{1}{\sqrt{1-r}} \times exp \left[ - \frac{Z^2}{2} \times r \right]
\label{abf}
\end{equation}

\noindent where $Z=\hat{\beta}/\sqrt{V}$ is the usual Z statistic and the shrinkage factor r is the ratio of the variance of the prior and total variance ($r = W/(V+W)$). 

\noindent Assuming a normal distribution, the $p$-value of each SNP can be converted to standard one-tailed Z-score by using inverse normal cumulative distribution function. So for a SNP, all that it is needed are the $p$-values from a standard regression output, and $\sqrt{W}$, the standard deviation of the normal prior N(0,W) on $\beta$. \\
The variance of the effect estimate, V, can be approximated using the MAF and sample size. However for imputed data it is preferable to use the variance outputted in standard regression analysis directly in the ABF equation.
For the expression dataset used here, the variance and effect estimates from the regression analysis were used for computation of ABFs. 
More details can be found in the Supplementary materials.

\subsection*{Choice of priors}  

\noindent Prior probabilities are assigned at the SNP level and correspond to mutually exclusive events. 
We assigned a prior of $1 \times 10^{-4}$ for $p_1$ and $p_2$, the probability that a SNP is associated with either of the two traits. Since all SNPs are assumed to have the same prior probability of association, this prior can be interpreted as an estimate for the proportion of SNPs that we expect to be associated with the trait in question. \\
We also assigned a prior probability of $1 \times 10^{-6}$ for $p_{12}$, the probability that one SNP is associated with both traits. This probability can be better understood when it is re-expressed as the conditional probability of a SNP being associated with trait 2, given that it is associated with trait 1. So assigning a probability of  $1 \times 10^{-6}$ means that 1 in 100 SNPs that are associated with trait 1 is also associated with the other. 
As a sensitivity analysis, we ran the comparison with Teslovich et al. using two other prior probabilities for $p_{12}$, $p_{12} =2 \times 10^{-6}$ which means 1 in 50 SNPs that are associated with one trait is also associated with the other; and $p_{12} =10^{-5}$ which means 1 in 10 SNPs. 

\noindent To compute the ABF, we also needed to specify the standard deviation for the prior, and we set this to 0.20 for binary traits and 0.15 for quantitative traits (more details in Supplementary).

\section*{Acknowledgments}

Claudia Giambartolomei is supported by a grant from the British Heart Foundation (CD6H). \\
The Diabetes and Inflammation Laboratory is funded by the JDRF, the Wellcome Trust (091157) and the National Institute for Health Research (NIHR) Cambridge Biomedical Research Centre. The Cambridge Institute for Medical Research (CIMR) is in receipt of a Wellcome Trust Strategic Award (100140).

\clearpage

\section*{Figures}

\clearpage

\section*{Tables}

\begin{table}[!ht]
 \caption{
  \bf{Loci previously reported to colocalise with liver eQTL, but not supported by our analysis}}
 \resizebox{\linewidth}{!}{%
    \begin{tabular}{llllllll cc | cc | l}  
      \hline\hline 
      Chr & Region & Gene &  Trait & Biom pval & Biom SNP  & eQTL pval & eQTL SNP &  \multicolumn{2}{c}{Primary signal} & \multicolumn{2}{c}{Secondary signal*} & Other genes colocalising in region (PP4 $>$ 75\%) \\ [0.5ex] 
      \hline 
          { }  & { }  & { } & { } & { } & { } & { } & { } & PP3 (\%)   & PP4 (\%) & PP4 (\%) & conditional SNP \\
          1 & 109824678:110224737 & {\it SYPL2} & LDL & 9.7e-171 & rs629301 & 7.1e-103 & rs2359653 & $>99$ & $<1$ & 99 &  rs2359653 &  {\it SORT1/CELSR2/PSRC1/PSMA5} \\  
          &  &  & TC & 8.0e-52  & rs672569 & 7.1e-103 & rs2359653 & $>99$ & $<1$ & 99 &  rs2359653 &  {\it SORT1/CELSR2/PSRC1/PSMA5} \\ \hline
          2 & 27467244:27867303 & {\it IFT172} & TG & 5.7e-133  & rs1260326 & 1.7e-130 & rs704791 & $>99$ & $<1$ &   &   &   {\it C2orf16/GCKR}  \\ 
          &  &  & TC & 7.3e-27 & rs1260326 & 1.7e-130 & rs704791 & $>99$ & $<1$ &   &  &  {\it C2orf16/GCKR} \\  \hline
          6 & 116062804:116462863 & {\it FRK} & LDL & 2.9e-09 & rs11153594 & 6.6e-15 & rs195517 & 99 & 1 &   &   \\
          &  &  &  TC & 1.7e-10 & rs9488822 & 6.6e-15 & rs195517 & 94 & 6 &   &   \\ \hline
         17 & 45589357:45989416 & {\it TBKBP1} & LDL & 1.1e-07 & rs8072100 & 2.1e-21 & rs9913503 & 87 & 9 &   &  & {\it KPNB1}  \\ 
          &  &  &   TC & 1.8e-07 & rs8072100 & 2.1e-21 & rs9913503 & 92 & 2 &   &   & {\it KPNB1} \\ \hline
         19 & 45248464:45648523 & {\it APOC4} & TG & 1.1e-30 & rs439401 & 1.1e-299 & 19:45452692:A\_AG & $>99$ & $<1$ & 96 & 19:45452692:A\_AG \\ \hline
         20 & 34013995:34414054 & {\it CPNE1} & TC & 3.8e-10 & rs2277862 & 7.3e-110 & rs6060524 & $>99$ & $<1$ &   &   \\ [1ex] 
   \hline 
    \end{tabular} }
    \begin{flushleft}Gene/eQTL associations previously reported as having a probable shared variant but not supported by our method based on PP3 (posterior probability for distinct signal values) $>75\%$. *Secondary signals are reported only when there is a secondary eQTL at a $p$-value greater than $10^{-4}$. Colocalisation tests are computed using the expression data conditioned on the listed SNP. Other genes in the same region as the gene listed that colocalise using our method are reported.
   \end{flushleft}
   \label{compare.tesl}
\end{table}

\begin{table}[!ht]
\caption{
\bf{Novel loci not previously reported to colocalise with liver eQTL, but colocalising based on our analysis}}
 \resizebox{\linewidth}{!}{%
\begin{tabular}{lllllll}
      \hline\hline 
Chr  &  Region  &  Gene  &   Trait  &  PP.H3.abf  &  PP.H4.abf & Reference \\ [0.5ex] 
      \hline 
2 &  20201795:20601854  &  {\it SDC1} &  TC  & 17 & 82  & \cite{Yilmaz2012} \\
2 &  85349026:85749085  &  {\it TGOLN2}  &  HDL  & 17 & 83 & \cite{Garver2002} \\
2 &  120908798:121308857  &  {\it INHBB}  &  LDL  & 7 & 77 & \cite{Johnson2012} \\
3 &  32322873:32722932  &  {\it CMTM6}  &  TC  & 8 & 77 \\
6 &  34355095:34755154  &  {\it C6orf106}  &  TC  & 15 & 85 \\
8 &  59158506:59558565  &  {\it UBXN2B}  &  LDL  & 13 & 87 & \cite{Wang2012}\\
   &    &    &  TC  & 15 & 85 \\
9 &  2454062:2854121  &  {\it VLDLR} &  LDL  & 1 & 91 & \cite{Nasarre2012} \\
10 &  17079389:17479448  &  {\it VIM}  &  TC  & 5 & 93 & \cite{Sarria1992} \\
10 &  94637063:95037122  &  {\it CYP26A1}  &  TG  & 3 & 95 & \cite{Loudig2005} \\
12 &  111508189:111908248  &  {\it CUX2}  &  HDL  & 2 & 89 \\
  &    &    &  LDL  & 2 & 98 \\
  &    &    &  TC  & 2 & 98 \\
15 &  96517293:96917352  &  ENSG00000259359  &  HDL  & 2 & 87 \\
16 &  56310220:56710279  &  {\it OGFOD1}  &  TC  & 7 & 84 & \cite{Saito2010} \\
16 &  71894416:72310900  &  {\it HP}  &  LDL  & 1 & 97 & \cite{Wassell1999} \\
  &    &    &  TC  & 1 & 97 \\
  &    &    &  TG  & 2 & 75 \\
  &    &  {\it HPR}  &  LDL  & 1 & 99 &  \cite{Nielsen2006} \\
  &    &    &  TC  & 1 & 99 \\
  &    &    &  TG  & 2 & 89 \\
22 &  46433083:46833138  &  {\it PPARA}  &  TC  & 10 & 81 & \cite{Staels2008} \\
  \hline 
    \end{tabular} }
    \begin{flushleft} Signals not reported by Teslovich et al. with PP4 $> 75\%$ for colocalisation between the liver eQTL dataset and the Teslovich meta-analysis of LDL, HDL, TG, TC, using the strict prior $p_{12}=10^{-6}$.
      For 11 genes with strong candidate status for lipid metabolism, we list a key reference that describes their function.
\end{flushleft}
  \label{novel}
  \end{table}

\clearpage

\bibliography{draft1_ref}

\end{document}


\beginsupplement

\begin{flushleft}
{\Large
\textbf{Bayesian Test for Colocalisation Between Pairs of Genetic Association Studies Using Summary Statistics}
}
\\
Claudia Giambartolomei$^{1,\ast}$, 
Damjan Vukcevic$^{2}$,
Eric E. Schadt$^{3}$,
Lude Franke$^{4}$,
Aroon D. Hingorani$^{5}$, 
Chris Wallace$^{6}$,
Vincent Plagnol$^{1}$

\bf{1} UCL Genetics Institute, University College London (UCL), Darwin Building, Gower Street, London WC1E 6BT, UK
\\
\bf{2} Murdoch Childrens Research Institute, Royal Children's Hospital, Melbourne, Australia
\\
\bf{3} Department of Genetics and Genomics Sciences, Mount Sinai School of Medicine, New York, New York, United States of America
\\
\bf{4} Department of Genetics, University Medical Center Groningen, University of Groningen, Groningen, The Netherlands 
\\
\bf{5} Institute of Cardiovascular Science, University College London, London WC1E 6BT, UK
\\
\bf{6} JDRF/Wellcome Trust Diabetes and Inflammation Laboratory, Cambridge, Institute for Medical Research, Department of Medical Genetics, NIHR, Cambridge Biomedical Research Centre, University of Cambridge, Addenbrooke's Hospital, Hills Rd, Cambridge, CB2 0XY, UK
\\ 
$\ast$ E-mail: claudia.giambartolomei.10@ucl.ac.uk
\end{flushleft}

\section*{Supplementary material}

\subsection*{Simplified model under assumption that all SNPs have the same prior}

The probability space for a single SNP can be fully partitioned into ($p_0, p_1, p_2, p_{12}$), where $p_0$ is the prior probability that a SNP is not associated with either trait, $p_1$ is defined as the prior probability that the SNP is only associated with trait 1, $p_2$ the prior probability that the SNP is associated only with trait 2, while $p_{12}$ is the prior probability that the SNP is associated with both traits. We therefore have:
$$p_0 + p_1 + p_2 + p_{12} = 1$$ 

\noindent The priors $p_0$, $p_1$, $p_2$, $p_{12}$ can vary across SNPs, for example depending on MAF, a measure of imputation quality, proximity to the promoter, function of the SNP. If however the priors do not vary across SNPs, then equation 1 in the main text becomes: 
\begin{equation}
L(H_h) = L(H_h \mid D) =\sum_{S \in S_h} P(D \mid S)P(S)= P(S \mid S \in S_h) \times \sum_{S \in S_h} P(D \mid S)
\label{simplified.lik}
\end{equation}

\noindent This simplification is possible because each binary vector that belongs to the same set $S_h$ has the same prior probability.

\noindent The prior probability of any one configuration in the different sets is: 
\begin{itemize} 
\item If $S \in S_0$, then $P(S) = p_0^Q$
\item If $S \in S_1$, then $P(S) = p_0^{Q-1} \times p_1$
\item If $S \in S_2$, then $P(S) = p_0^{Q-1} \times p_2$
\item If $S \in S_3$, then $P(S) = p_0^{Q-2} \times p_1 \times p_2$
\item If $S \in S_4$, then $P(S) = p_0^{Q-1} \times p_{12}$
\end{itemize} 

\noindent Because we condition our analysis on having at most one association per trait, these probabilities should be normalised: $P(S) = p_0^Q /C$, where $C$ is the sum of the probabilities associated with all the assignments that contain at most one association per trait.
However, because subsequent derivations only consider the ratio of probabilities, the $C$ term cancels out and this normalisation becomes unnecessary.
Dividing each of these probabilities by $P(S_0)$ to obtain the ratio of prior odds, and since $p_0 \approx 1$, the second terms in equation 4 in the main text becomes: 
\begin{itemize} 
\item If $S \in S_0$, then $\frac{P(S)}{P(S_0)} = \frac{p_0^Q}{p_0^Q} =1 $
\item If $S \in S_1$, then $\frac{P(S)}{P(S_0)} = \frac{p_0^{Q-1}} {p_0^Q} \times p_1 = \frac{p_1} {p_0} \approx p_1$
\item If $S \in S_2$, then $\frac{P(S)}{P(S_0)} = \frac{p_0^{Q-1}} {p_0^Q} \times p_2 = \frac{p_2} {p_0} \approx p_2$
\item If $S \in S_3$, then $\frac{P(S)}{P(S_0)} = \frac{p_0^{Q-2}} {p_0^Q} \times p_1 \times p_2 = \frac{p_1} {p_0} \times \frac{p_2} {p_0} \approx p_1 \times p_2$
\item If $S \in S_4$, then $\frac{P(S)}{P(S_0)} = \frac{p_0^{Q-1}} {p_0^Q} \times p_{12} = \frac{p_{12}} {p_0} \approx p_{12}$
\end{itemize} 

\bigskip 
\noindent To compute the first terms in equation 4 in the main text, the BFs for each configuration in a set, we make use of the ABF derived for each SNP-trait association (section below). Two key assumptions are necessary for the following computations. Firstly that the traits are measured in unrelated individuals, and secondly that the effect sizes for the two traits are independent. 

Putting the two terms together, we have:

\begin{itemize}
\item $ L(H_0) \slash L(H_0) = 1 $
\item $ L(H_1) \slash L(H_0) = p_1 \times \sum_{j=1}^{Q} ABF_j^1 $
\item $ L(H_2) \slash L(H_0) = p_2 \times \sum_{j=1}^{Q} ABF_j^2 $
\item $ L(H_3) \slash L(H_0) = p_1 \times p_2 \times \sum_{j, k, j \neq k} ABF_j^1 ABF_k^2 $ 
\item $ L(H_4) \slash L(H_0) = p_{12} \times \sum_{j=1}^Q ABF_j^1 \times ABF_j^2 $ 
\end{itemize}

\bigskip 

\noindent Of note, we can also write:
$$  L(H_3) \slash L(H_0) =    p_1 \times p_2 \times \sum_{j=1}^{Q} ABF_j^1 \sum_{j=1}^{Q} ABF_j^2 - \left[ \frac{p_1 \times p_2}{p_{12}} \times L(H_4) \slash L(H_0) \right]$$

\subsection*{Bayes factor computation}
We assume that summary statistics for each SNP in the two datasets were obtained by fitting a generalised linear model with the phenotype as dependent variable and SNP genotype call as independent variable:
$$Y = \mu + \beta X$$
The Bayes factor quantities are estimated from summary statistics using the Asymptotic Bayes Factor derivation \cite{Wakefield2009}. 
Using Wakefield's notations, under the null we assume that the effect size $\beta = 0$.
Under the alternative, $\beta$ is normally distributed with mean $0$ and variance $W$.

To derive the ABF computation, Wakefield uses the fact that asymptotically $\hat{\beta} \to N(\beta, V) $.
The distribution of the estimated regression parameters $\hat{\mu}$ and $\hat{\beta}$ is:

\begin{equation}
\left[
\begin{array}{l}
\hat{\mu} \\
\hat{\beta}
\end{array}
\right] \sim N_{p+1} \left(
\left[
\begin{array}{l}
\mu \\
\beta
\end{array}
\right]
\left[
\begin{array}{ll}
\mathbb{I}_{\mu \mu} & \mathbb{I}_{\mu \beta} \\
\mathbb{I}_{\mu \beta}^T & \mathbb{I}_{\beta \beta}
\end{array}
\right]^{-1}
\right)
\end{equation}

\noindent The intercept $\mu$ is a nuisance parameter which we can remove using the transformation:
$$ \gamma = \mu +  \frac{\mathbb{I}_{\mu \beta}}{\mathbb{I}_{\mu \mu}} \beta $$
In which case the previous equality becomes:
 
\begin{equation}
\left[
\begin{array}{l}
\hat{\gamma} \\
\hat{\beta}
\end{array}
\right] \sim N_{p+1} \left(
\left[
\begin{array}{l}
\gamma \\
\beta
\end{array}
\right]
\left[
\begin{array}{ll}
\mathbb{I^*}_{\mu\mu} & 0 \\
0^T & \mathbb{I}_{\beta\beta}
\end{array}
\right]^{-1}
\right)
\end{equation}

\noindent This independence property can be combined with independent priors for both parameters so that one can only consider $\beta$ and ignore the effect of the $\mu$ term.
Hence, after applying the reparameterisation we have:
$$ BF= \int \frac{f(\beta)}{f(0)} \pi(\beta) d\beta$$
\noindent Which then becomes (assuming normality and asymptotic behaviour):
$$ABF = \frac{1}{\sqrt{1-r}} \times exp \left[ - \frac{Z^2}{2} \times r \right] $$ \label{abf}
where $Z = \hat{\beta} / \sqrt{V}$ is the Wald test statistic.
The shrinkage factor r is the ratio of the variance of the prior and total variance ($r = W/(V+W)$). 
This ratio takes a value between 0 and 1 and measure the relative contributions of the prior (W) and likelihood (V) to the inference. Values of r closer to 1 indicate a larger contribution from the likelihood (i.e. the data). The asymptotic posterior distribution of $\beta$ is $N(r\hat{\beta}, rV)$. As the sample size increases, V $\to$ 0 and r $\to$ 1, so that the posterior concentrates around the MLE. 
We can compute the Z statistic from the $p$-values of a standard regression output using $|z| = \Phi^{-1} (1-p/2)$, where $\Phi^{-1}$ is the inverse normal cumulative. Otherwise we can compute the Wald statistics directly if the estimated regression coefficients $\hat{\beta}$ and their variances $var(\hat{\beta})$ are supplied.

\subsection*{Variance of the estimated effect size $Var(\hat{\beta}) = V$}
The variance of the maximum likelihood estimate $\hat{\beta}$ (denoted by $V$) can be approximated using the allele frequency of the variant $f$, the sample size $N$ and the case control ratio  $s$ for binary outcome.
It is well established that the score statistic to test whether the effect size $\beta = 0$ is:
$$U = \sum_i (Y_i - \bar{Y}) X_j$$
Score test theory asserts that the variance of $U$ under the null is the inverse of the variance of the estimated effect size $\hat{\beta}$, also under the null.
$Var(U)$ under the null can be estimated in several ways, including the standard derivation of the Fisher information matrix \cite{McCullagh1983}.

\noindent We assume that the genotypes $X$ are under Hardy Weinberg equilibrium.
Hence $X$ is drawn from a binomial distribution with success parameter f, which we use to denote the allele frequency.
$$Var[X_j] = 2 f_j (1-f_j)$$
where $f_j$ is the population allele frequency of the SNP j. 

When the dataset is a case-control (the trait Y is binomial):
$$Var[Y] = s (1-s)$$
where $s$ is the proportion of cases in the population. 
If the outcome variable $Y$ is continuous we assume that $Y$ is normalised such that $Var(Y) = 1$.

\noindent Putting together these equations, we have, in a case control setting:
$$  V = Var(\hat{\beta}_j) = \frac{1}{N s(1-s) \times 2f_j(1-f_j)} $$


\subsection*{Choice of priors for the probability that each variant affects the traits}

The prior probabilities assigned to each SNP, $p_0, p_1, p_2, p_{12}$, are mutually exclusive events. 
The probability of a SNP being associated with both traits, $p_{12}$, can be interpreted using a conditional argument:

\begin{eqnarray*}
  p_{12} & = &   \mathbb{P}(\textrm{SNP associated with both traits})\\
  & = & \mathbb{P}(\textrm{SNP associated with trait 1}) \times  \mathbb{P}(\textrm{SNP associated with trait 2} \quad |  \quad \textrm{SNP associated with trait 1})
\end{eqnarray*}

The conditional term in the right hand side of the equation above can be approximated as follows:

\begin{eqnarray*}
  \mathbb{P}(\textrm{SNP associated with trait 2} \quad |  \quad \textrm{SNP associated with trait 1}) & = &  \frac{p_{12}}{(p_{12} + p_1)} \\
  & = &  \frac{10^{-6}} {(10^{-6} + 10^{-4})} \\
  & \approx & \frac{10^{-6}}{10^{-4}}  \\
  & \approx &  0.01
\end{eqnarray*}

\noindent So in terms of conditional probability, if we assume a prior probability of $p_{12} = 1 \times 10^{-6}$ and $p_1 = p_2 = 1 \times 10^{-4}$, then the prior assumption is that of all SNPs associated with trait 1, 1 in 100 of them will also be associated with trait 2. Since as we stated previously each SNP belongs to only one of the five sets corresponding to the five hypotheses, the probability of a SNP not being associated with either trait is:
$p_0 = 1 - (p_1 + p_2 + p_{12}) \approx 0.9998 \approx 1$.

\subsection*{Choice of priors for the standard deviation W of the effect size parameter $\beta$}
Prior standard deviation of the additive effect parameter $\beta$ was set to 0.15 for a continuous trait. Owing to our assumption that $Var(Y) = 1$, this prior corresponds to a variance explained of $\sim$ 0.01 for a variance with MAF of $30\%$. 

In a case-control study, we set $W = 0.2$ for the variance of the log-odds ratio parameter, as was previously used in WTCCC \cite{Burton2007}, which corresponds to an inter-quartile range between 0.74 and 1.35 for the odds ratio parameter.
The priors chosen for case-control and for quantitative traits are also very similar to the SNPTEST default (\cite{Marchini2007, Marchini2010}).


\subsection*{Simulation procedure}
To simulate the posterior probability of a common signal (``PP4") under different scenarios, we used the imputed genotypes from two different datasets: Whitehall II study (WHII), a longitudinal prospective cohort study genotyped using the gene-centric Illumina Metabochip \cite{Voight2012, Talmud2009}, and the expression dataset described herein. 
We randomly chose a causal SNP A among genotyped and well imputed SNPs (Rsq $>$ 0.8) from any genomic region in common between the two datasets. 

The additive genetic variance explained by the locus is
$$Var[\beta X_A]= \beta_A^2 \times 2f_A (1-f_A)$$ \label{var.expl}
where $f_A$ is the population allele frequency of the causal SNP A, and $\beta_A$ is the additive effect (in standard deviations) \cite{Falconer}. 

For different variance explained for the causal SNP , depending on the simulation scenario, we computed the true effect at the causal SNP:
$$\beta_A = \sqrt{\frac{Var[\beta X_A]} {2f_A (1-f_A)}}$$ \label{true.effect}

We then simulated the phenotype of the $i^{th}$ individual in each of the two datasets using the computed true effect $\beta_A$: 
$$Y_i = x_{iA} \times \beta_A + e_i \quad \quad with \quad e_i \sim N(0, 1)$$ \label{sim.pheno}
where $x_{iA}$ is the additive genetic value at the causal SNP of individual $i$ and $e_i$ is a random error drawn from a normal distribution with mean 0 and variance of 1.0. 
  
To simulate $p$-values from different sample sizes using the original datasets, we computed the expected value of each estimated beta multiplying the pairwise correlation coefficient r between the causal SNP A and all other SNPs, by the true value of beta at the causal SNP:

\noindent The derivation of this equation is described in the following section.

\noindent The expected regression coefficient at a second SNP B in relation to the causal SNP A is then:
\begin{equation}
  \beta_j  =  \beta_A \, r \sqrt\frac{f_A (1-f_A)}
                                     {f_j (1-f_j)}  \, .
\label{expected.effect}
\end{equation}

where $f_j$ is a vector or $\beta$s for all SNPs excluding the causal SNP.

\noindent The difference between the regression coefficients $\hat{\beta}$ estimated from the glm and the expected value $\beta$, computed at each SNP, can be considered the standard error of the mean and it varies with $\frac{1}{\sqrt(N)}$, where N is the original sample size. When N is increased, the estimate becomes closer to the true value $\beta$, decreasing the variability of the estimator. Then the simulations with a larger sample size involves simply rescaling this difference:

\begin{equation}
(\hat{\beta} - \beta) \sqrt{\frac{N} {N.new}}
\label{diff.effect}
\end{equation}

\noindent Then we can compute the new $p$-values from the new simulated estimates and standard deviations.
We used this method to perform simulations to find sample size required for colocalisation analysis, and to find the consequence of using limited variant density. 

 
\subsection*{Relationship between model parameters and LD}
Equation \ref{expected.effect} was first derived in \cite{Vukcevic2011}, where the additive effects for case-control studies were shown to decay linearly, in proportion to $r$, the correlation between the causal and marker loci. Here we show the same relationship holds for quantitative traits. 

\noindent We use the same LD model, and associated notation, as defined in \cite{Vukcevic2011}.  To summarise briefly, let $A$ and $B$ be a pair of biallelic SNPs, with the alleles at each coded by $0$ and $1$.  Let $f_A$ be the population frequency of allele $1$ at SNP~$A$, and define $f_B$ similarly for SNP~$B$\@.  Let $r$ be the population correlation coefficient between them, the square of which is a commonly used measure of LD\@. We can define the following conditional probabilities on a haplotype level:

\[
  \begin{array}{cc}
    q_0 = P(A=1 \mid B=0)  \, ,  \\
    q_1 = P(A=1 \mid B=1)  \, ,  \\
  \end{array}
\]

\noindent These quantities were shown to be related by the identity,
\[
  r  =  \left(q_1 - q_0\right) \sqrt\frac{f_B (1 - f_B)}
                                         {f_A (1 - f_A)}  \, ,
\]

\noindent Let SNP~$A$ be a causal and SNP~$B$ be a marker.  Define the following expectations:
\[
  \begin{array}{cc}
    a_0 = \E(Y \mid A=0)  \, ,  & \qquad  b_0 = \E(Y \mid B=0)  \, ,  \\
    a_1 = \E(Y \mid A=1)  \, ,  & \qquad  b_1 = \E(Y \mid B=1)  \, ,  \\
    a_2 = \E(Y \mid A=2)  \, ,  & \qquad  b_2 = \E(Y \mid B=2)  \, .
  \end{array}
\]
Relating these using the LD model gives,
\begin{align*}
  b_0  &=  a_0 (1-q_0)^2      + a_1 2 q_0 (1-q_0) + a_2 q_0^2    \, ,  \\
  b_1  &=  a_0 (1-q_0)(1-q_1) + a_1 \left(q_0 (1-q_1) +
                                          q_1 (1-q_0)\right)
                                                  + a_2 q_0 q_1  \, ,  \\
  b_2  &=  a_0 (1-q_1)^2      + a_1 2 q_1 (1-q_1) + a_2 q_1^2    \, .
\end{align*}
Combining and rearranging these gives,
\[
  b_1^2 - b_0 b_2  =  \left(a_1^2 - a_0 a_2\right)
                      \left(q_1   - q_0    \right)^2  \, .
\]
If we consider an additive model, it is easy to show that $b_1^2 - b_0 b_2 =
\beta_B^2$ and $a_1^2 - a_0 a_2 = \beta_A^2$.  Putting these together gives,
\begin{equation}
  \beta_B  =  \beta_A \, r \sqrt\frac{f_A (1 - f_A)}
                                     {f_B (1 - f_B)}  \, .
\label{eqn:beta.versus.r}
\end{equation}

Equation \ref{eqn:beta.versus.r} is analogous to the respective results from \cite{Vukcevic 2011} for binary traits. The main difference is that for quantitative traits here we have shown them to be exact.  As before, we can see that the deviation effect decays more quickly with LD than does the additive effect: quadratically in $r$ rather than linearly. This implies that the distortion effect described earlier for binary traits will extend also to quantitative traits.

\section*{Relationship between model parameters and LD using imputed genotypes}

The previous section assumed the genotypes were observed without error.  We can
derive a similar relationship in the scenario where we instead use imputed
genotype data.

Continuing with the same notation, we now allow $A$ and $B$ to have posterior
distributions for each individual's genotype.  The appropriate analysis should
average the likelihood over these posteriors.  A convenient approximation to
this is to replace each genotype by its mean posterior value.
\cite{Guan2008} showed that this is a good approximation in the GWAS context.
This leaves us in a similar situation as before, except that $A$ and $B$ have
become continuous quantitative variables.  The following general results apply
for the regression parameters,
\begin{align*}
  \beta_A &= \frac{\cov(Y, A)}{\var(A)}  \, ,  \\
  \beta_B &= \frac{\cov(Y, B)}{\var(B)}  \, .
\end{align*}

We need two assumptions to complete the derivation.  Firstly, we require that
$Y$ and $B$ are conditionally independent given $A$.  In other words, $\cov(Y,
B \mid A) = 0$.  This is implicit in the definition of SNP $A$ being the
\emph{causal} SNP, and is also implicitly assumed in the definitions in the
previous section.  The only extension here is that we assume this relationship
still holds after imputation.  This allows us to simplify the covariance of $Y$
and $B$ using the law of total covariance,
\begin{align*}
     \cov(Y, B)
  &= \E(\cov(Y, B \mid A)) + \cov(\E(Y \mid A), \E(B \mid A))  \\
  &= \E(0)                 + \cov(\E(Y \mid A), \E(B \mid A))  \\
  &= \cov(\E(Y \mid A), \E(B \mid A))  \, .
\end{align*}
The second assumption we need is that the conditional expectations of $Y$ and
$B$ on $A$ are both linear in $A$.  That is,
\begin{align*}
  \E(Y \mid A) &= \theta_{Y0} + \theta_{Y1} A  \, ,  \\
  \E(B \mid A) &= \theta_{B0} + \theta_{B1} A  \, .
\end{align*}
We already make this assumption for the relationship between $Y$ and $A$, in
using a linear regression model for the trait.  The extra part here is that we
assume the same type of relationship between the causal SNP and marker SNP.
Since the underlying values for $A$ and $B$ are binary, this is reasonable.

Note that these are just standard linear regression equations, so we have that,
\begin{align*}
  \theta_{Y1} &= \frac{\cov(Y, A)}{\var(A)}  \, ,  \\
  \theta_{B1} &= \frac{\cov(B, A)}{\var(A)}  \, .
\end{align*}

Putting all of these formulae together gives,
\begin{align*}
     \cov(Y, B)
  &= \cov(\E(Y \mid A), \E(B \mid A))    \\
  &= \cov(\theta_{Y1} A, \theta_{B1} A)  \\
  &= \theta_{Y1} \theta_{B1} \cov(A, A)  \\
  &= \frac{\cov(Y, A) \cov(B, A)}{\var(A)}  \, .
\end{align*}
Dividing both sides by $\var(B)$ gives,
\begin{align*}
\frac{\cov(Y, B)}{\var(B)}
        &= \frac{\cov(Y, A)}{\var(A)}  \,
           \frac{\cov(B, A)}{\var(B)}  \\
\frac{\cov(Y, B)}{\var(B)}
        &= \frac{\cov(Y, A)}{\var(A)}                \,
           \frac{\cov(B, A)}{\sqrt{\var(A) \var(B)}} \,
           \sqrt{\frac{\var(A)}{\var(B)}}            \\
\beta_B &= \beta_A \cor(B, A) \sqrt{\frac{\var(A)}{\var(B)}}  \\
\beta_B &= \beta_A \, r       \sqrt{\frac{\var(A)}{\var(B)}}  \, .
\end{align*}
This formula is analogous to equation~\eqref{eqn:beta.versus.r}.  Note that if
$A$ and $B$ are observed without error then we recover the previous formula.

\subsection*{Simulations to find sample size required for colocalisation analysis}

We used the simulation method described above to compare the distribution of $PP4$ of randomly sampled regions under different scenarios. We used the original sample size of the eQTL dataset (n=966) and a variance explained of 10\% for expression, while varying the sample size of the biomarker dataset and the proportion of the biomarker's variance explained by the causal variant.

\subsection*{Simulations to find consequence of using limited variant density}

\begin{figure}[!ht]    
\begin{center}
\includegraphics[width=6in]{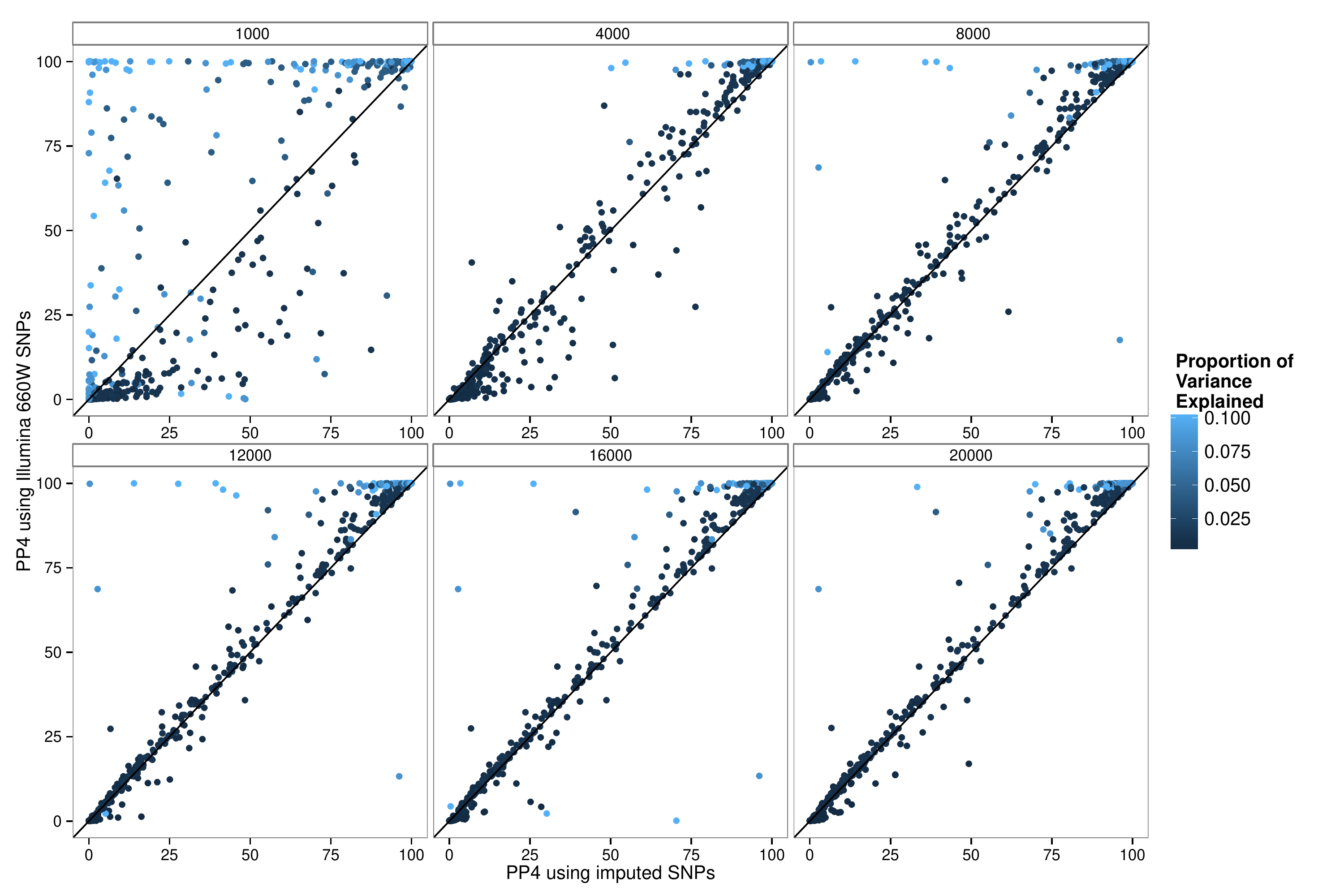}
\end{center}
\caption{
 {\bf Simulation analysis with a shared causal variant between two studies, one eQTL (sample size 966 samples) and one biomarker, comparing results using imputed versus not imputed data where the causal SNP is included in both the cases.} Each plot shows different sample sizes for one dataset. The variance explained by the causal variant for both the traits is colour coded.
    The x-axis shows the estimated PP4 for 1,000 simulations using data imputed from metaboChip Illumina array (Methods). 
    The y-axis uses the same dataset restricted to variants present on the Illumina 660W genotyping array to assess the impact of a lower variant density.
    The causal variant is included in the Illumina 660W panel.}  
\label{illumina_vs_meta_causal}
\end{figure}

\begin{figure}[!ht]    
\begin{center}
\includegraphics[width=6in]{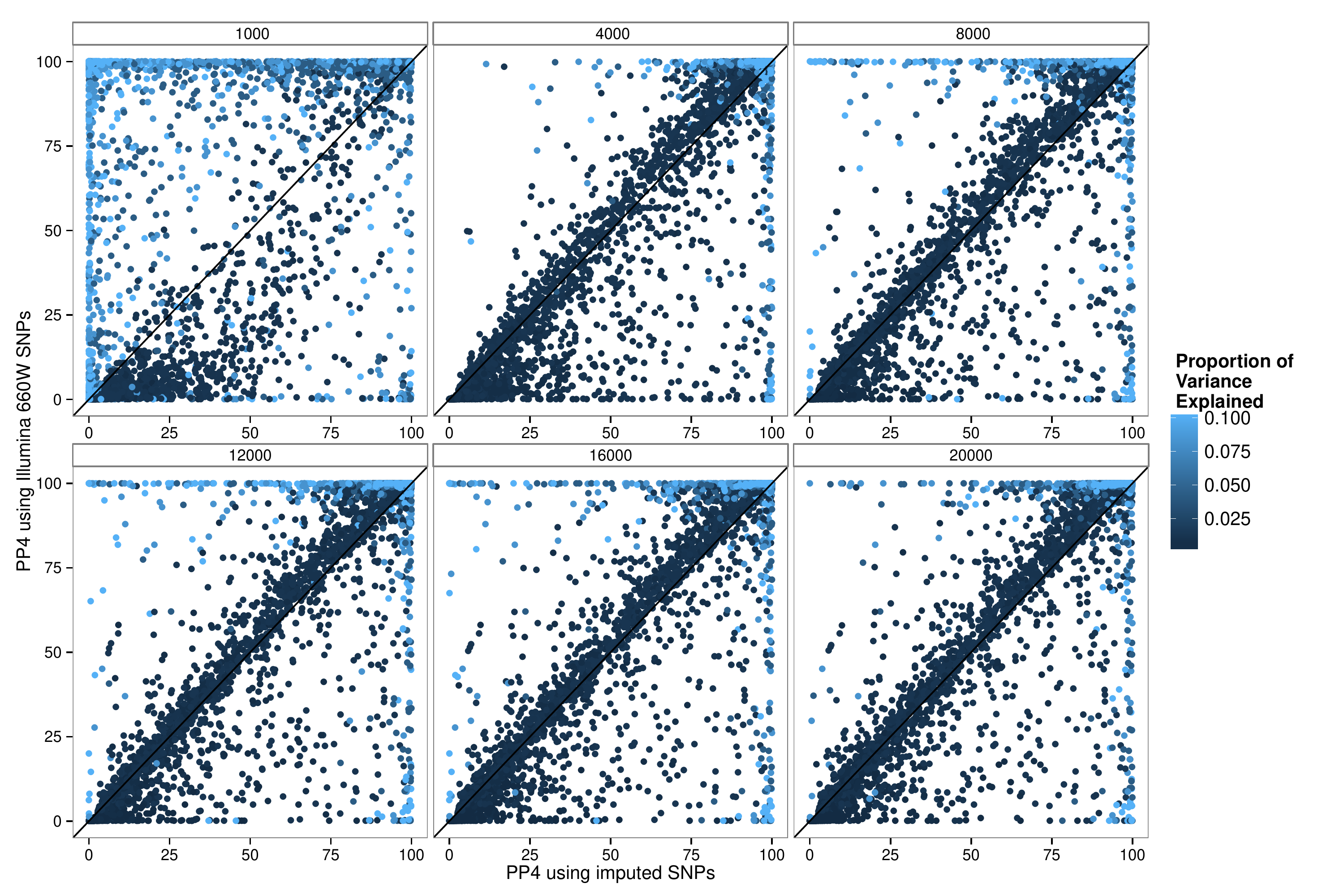}
\end{center}
\caption{
{\bf Simulation analysis with a shared causal variant between two studies, one eQTL (sample size 966 samples) and one biomarker, comparing results using imputed versus not imputed data where the causal SNP is not included in one of the datasets.} Each plot shows different sample sizes for one dataset. The variance explained by the causal variant for both the traits is colour coded. Column and row headings are the same as in previous figure. The causal SNP is not included in Illumina 660W panel.}  
\label{illumina_vs_meta_nocausal}
\end{figure}

We compared the $PP4$ of randomly sampled regions, using the simulation procedure described above, with the $PP4$ of the same regions after filtering for only the SNPs present in the Illumina 660K Chip (Figures \ref{illumina_vs_meta_causal} and \ref{illumina_vs_meta_nocausal}). The original dataset based on $>99$0genomes imputed data will be much denser than the Illumina dataset. This way we can consider the  consequences of having imputed data versus genotyped data. 

The same procedure is used to simulate the case when the causal SNP is not included in the data by excluding the causal SNP only from the Illumina dataset. 
All analyses were conducted in R \cite{R}.

\subsection*{Simulations for comparison with existing colocalisation tests}

\begin{figure}[!ht]  
  \begin{center}
    \includegraphics[width=6in]{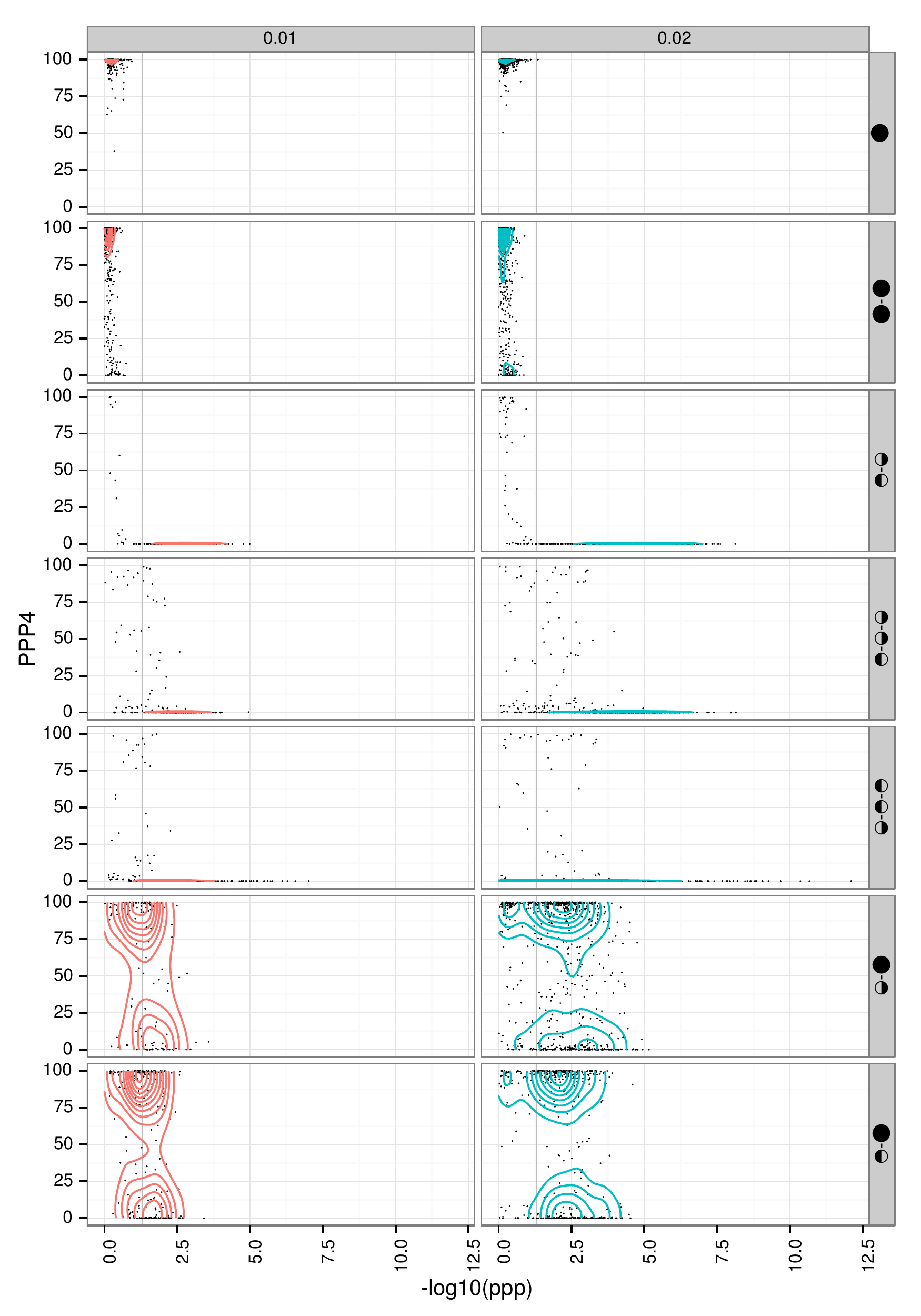}
  \end{center}
\caption{
 {\bf The relationship between PP4 and the posterior predictive $p$-value (on a -log10 scale) from proportional testing.} Proportional testing uses the BMA approach, integrating over all possible two SNP models.  Each row shows a different scenario, the total number of causal variants in a region is indicated by number of symbols in the plot titles with the type of causal variant indicated by the symbol: full circle - affects both traits; top only - affects one trait; bottom only- affects other trait.  For proportional testing, the grey vertical line indicates the threshold ppp of 0.05. Each column shows the total proportion of trait variance for the biomarker explained by all variants in a region, with variance explained spread equally over all variants.  In all cases, for the eQTL trait, n=1,000, 10\% of the variance explained by the variant; for the biomarker trait, n=10,000.}
    \label{compare_methods}
\end{figure}

\begin{figure}[!ht]  
  \begin{center}
    \includegraphics[width=6in]{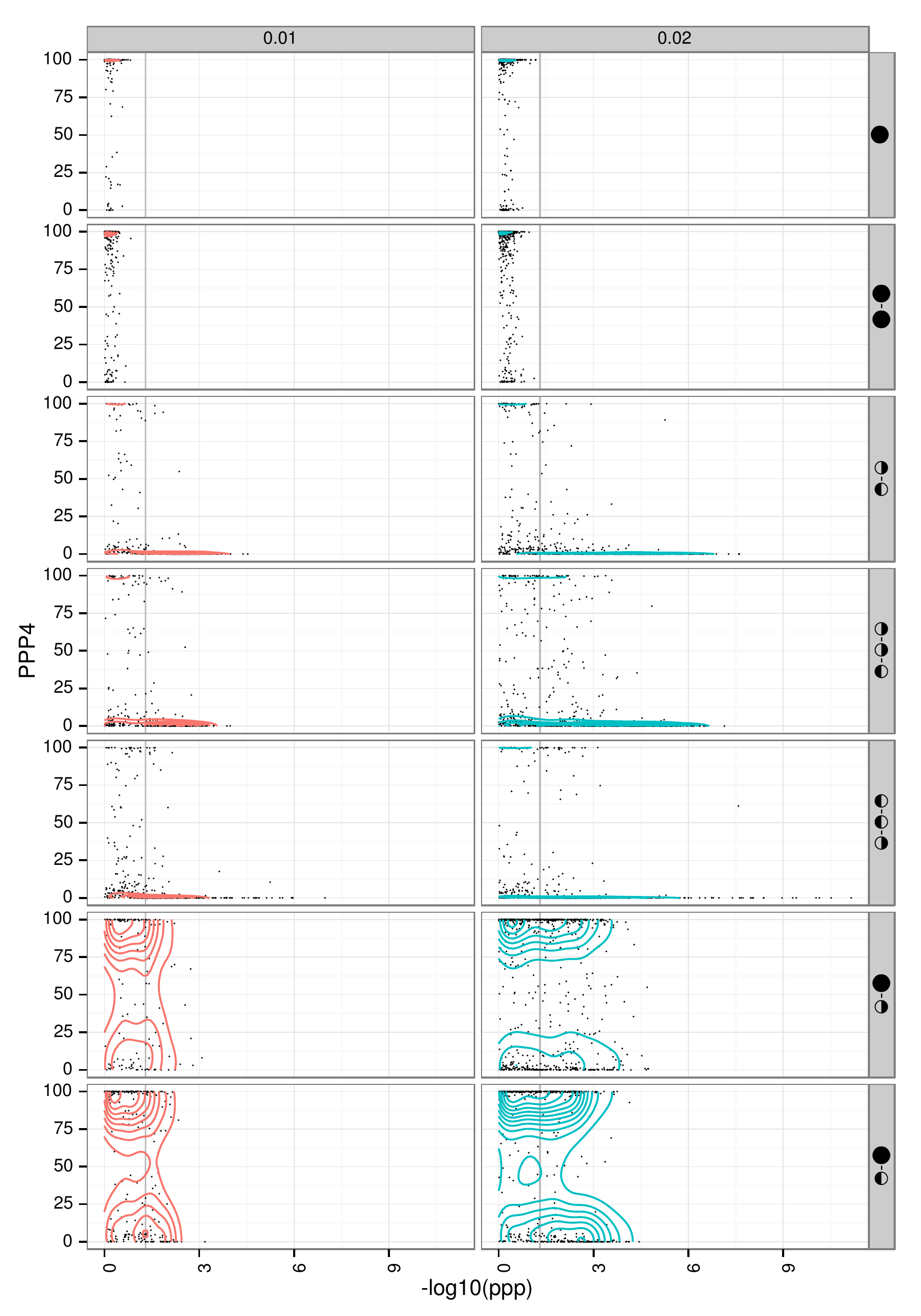}
  \end{center}
\caption{
 {\bf The relationship between PP4 and the posterior predictive $p$-value (on a -log10 scale) from proportional testing, using subset of SNPs which appear on the Illumina HumanOmniExpress genotyping array.} For the eQTL trait, n=1,000, 10\% of the variance explained by the variant; for the biomarker trait, n=10,000, 1\% or 2\% of the variance explained by the variant. Column and row headings are the same as in previous figure.}
    \label{compare_methods_chip}
\end{figure}

For simulations comparing proportional and Bayesian approaches,  we sampled, with replacement, haplotypes of SNPs with a minor allele frequency of at least 5\% found in phased $>99$0 Genomes Project data\cite{AutosomesChromosome2012} across all 49 genomic regions outside the major histocompatibility complex (MHC) which have been identified as type 1 diabetes (T1D) susceptibility loci to date, as summarised in T1DBase (Figures \ref{compare_methods} and \ref{compare_methods_chip}).
\cite{Burren2011}.  These represent a range of region sizes and genomic topography typical of GWAS hits.  We excluded the MHC region which is known to have high variation, strong LD and exhibits huge genetic influence on autoimmune disease risk involving multiple loci and hence requires individual treatment in any GWAS \cite{Nejentsev2007}.
For each trait, we selected one or two ``causal variants'' at random, and simulated a Gaussian distributed quantitative trait for which each causal variant SNP explains a specified proportion of the variance. We either used all SNPs or the subset of SNPs which appear on the Illumina HumanOmniExpress genotyping array to conduct colocalisation testing to reflect the scenarios of very dense targeted genotyping \emph{versus} a less dense GWAS chip.  All analyses were conducted in R \cite{R} using the \texttt{coloc} package for proportional colocalisation testing.

\clearpage

\subsection*{Supplementary Figures and Tables}

\newpage
\newgeometry{top = 1.5cm, bottom=0.1cm}
\restoregeometry
\clearpage

\newpage
\newgeometry{top = 1.5cm, bottom=0.1cm}
\restoregeometry


\newgeometry{top = 1.5cm, bottom=0.1cm}
\begin{figure}[htb]  
\centering
    \includegraphics[trim={5.5cm 2cm 6cm 4.5cm},clip]{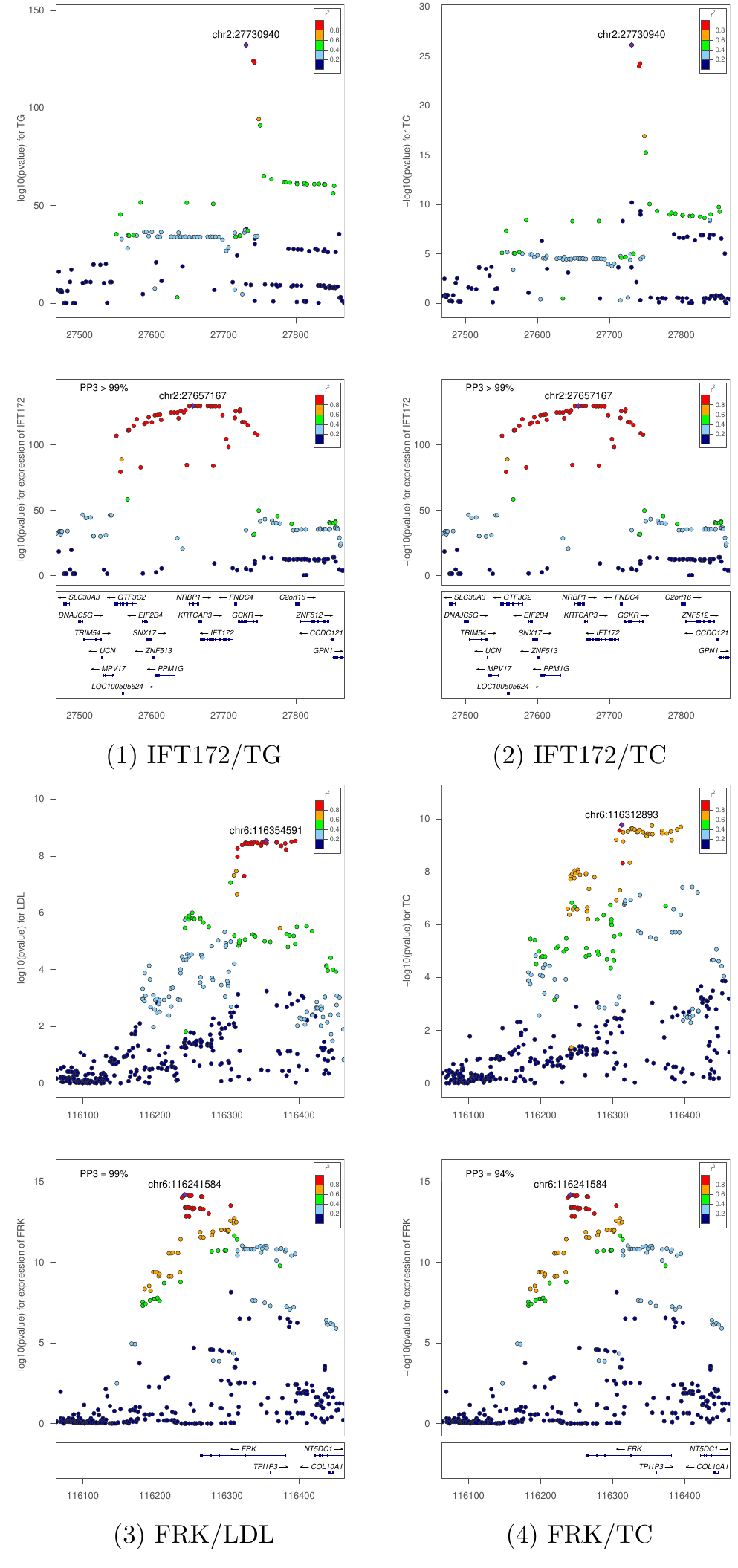}   \\%
  \caption*{}
\end{figure}
\clearpage

\begin{figure}[htb]
\centering
        \includegraphics[trim={5.5cm 2cm 6cm 4.5cm},clip]{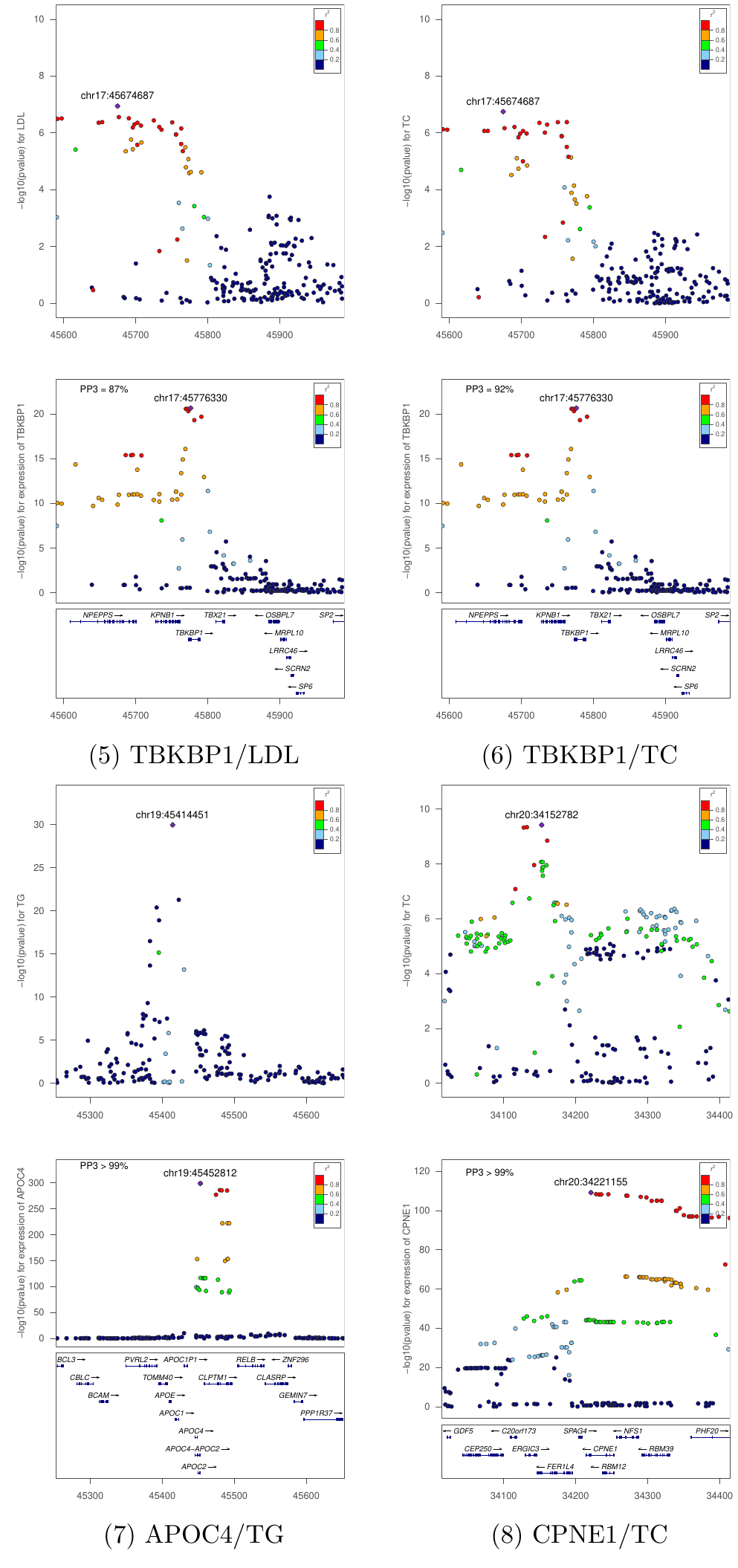} \\ 
\caption{
 {\bf  Regional Manhattan plots with $-log10$ $p$-value for loci reported in Teslovich et al. \cite{Teslovich2010} as having a probable shared variant but not supported by our method based on PP4 (posterior probability for a shared signal values) $\mathbf{< 75\%}$.} The plots focus on a specific region of the genome with a range of $\sim 400$ kilobases around the expression probe of the gene specified below each plot. The top plots use the -log10($p$-value) from the published meta-analysis with one of the four lipid biomarkers; the bottom plots show the -log10($p$-value) computed by fitting a generalized linear model with expression as dependent variable and SNP genotypes as independent variable. Each dot represents one SNP, imputed or directly typed. The value on the top of each plot shows the PP4 from the colocalisation test between the two top SNP of the expression and biomarker associations.}
 \label{compare_tesl_plots}
\end{figure}
\restoregeometry
\clearpage


\newpage
\begin{landscape}
\pagestyle{empty}
\fudge{3cm}{1.5cm}
\advance\LTcapwidth by 360pt 
\tiny  
\setlength\LTleft{-80pt}            
\setlength\LTright{-30pt}           
\begin{longtable}{@{\extracolsep{\fill}}l l l l l l l l l l l l | c c | c c@{}}
\caption{
\bf{Results using reported loci that colocalise with liver eQTL}}
\tabularnewline \hline
   geneTranscriptsymbol & TranscriptPvalueTesl & Trait & Chr & Region & Biom pval  &  Biom SNP   &  eQTL pval  &  eQTL SNP  &  Best Causal &\multicolumn{2}{c}{$p_{12}=10^{-5}$} & \multicolumn{2}{c}{$p_{12}=2\times 10^{-6}$} & \multicolumn{2}{c}{$p_{12}=10^{-6}$} \\ [0.5ex] 
\hline
 &  &  &  &  &  & & & & & PP3 (\%) & PP4 (\%) & PP3 (\%) & PP4 (\%) & PP3 (\%) & PP4 (\%)\\
RHCE & 7.00E-54 &   LDL,TC   &   NA   &   NA   & NA & NA & NA & NA & NA &   NA   &   NA   &   NA   &   NA   &   NA   &   NA   \\
RHD   & 4.00E-08 &   LDL   & 1 &   25428038:25828097   & 1.20E-10 & rs12027135 & 7.70E-11 & rs909832 & rs12027135 & 9 & 91 & 33 & 67 & 50 & 50 \\
   &      &      & 1 &   25456834:25856893   & 1.20E-10 & rs12027135 & 0.0087 & rs909832 & rs12027135 &   $<1$   & 2 &   $<1$   &   $<1$   &   $<1$   &   $<1$   \\
   &      &   TC   & 1 &   25428038:25828097   & 4.10E-11 & rs12027135 & 7.70E-11 & rs909832 & rs12027135 & 9 & 91 & 33 & 67 & 50 & 50 \\
   &      &      & 1 &   25456834:25856893   & 4.10E-11 & rs12027135 & 0.0087 & rs909832 & rs12027135 &   $<1$   & 2 &   $<1$   &   $<1$   &   $<1$   &   $<1$   \\
TMEM50A   & 4.00E-08 &   LDL   & 1 &   25488669:25888728   & 1.20E-10 & rs12027135 & 4.40E-11 & rs9689 & rs3091242 & 12 & 88 & 41 & 59 & 58 & 42 \\
   &      &   TC   & 1 &   25488669:25888728   & 4.10E-11 & rs12027135 & 4.40E-11 & rs9689 & rs3091242 & 13 & 87 & 43 & 57 & 60 & 39 \\
TMEM57   & 2.00E-145 &   LDL   & 1 &   25626305:26026364   & 1.20E-10 & rs12027135 & 2.10E-31 & rs873308 & rs12027135 & 1 & 99 & 5 & 94 & 10 & 90 \\
   &      &      & 1 &   25624780:26024839   & 1.20E-10 & rs12027135 & 6.40E-08 & rs686631 & rs10903129 & 2 & 98 & 9 & 90 & 17 & 82 \\
   &      &      & 1 &   25625807:26025866   & 1.20E-10 & rs12027135 & 7.60E-223 & rs873308 & rs873308 & 2 & 97 & 11 & 88 & 21 & 79 \\
   &      &   TC   & 1 &   25626305:26026364   & 4.10E-11 & rs12027135 & 2.10E-31 & rs873308 & rs12027135 & 1 & 99 & 5 & 94 & 11 & 89 \\
   &      &      & 1 &   25624780:26024839   & 4.10E-11 & rs12027135 & 6.40E-08 & rs686631 & rs12027135 & 2 & 98 & 9 & 90 & 17 & 82 \\
   &      &      & 1 &   25625807:26025866   & 4.10E-11 & rs12027135 & 7.60E-223 & rs873308 & rs873308 & 3 & 97 & 15 & 85 & 26 & 74 \\
ANGPTL3   & 1.00E-13 &   LDL   & 1 &   62870388:63270447   & 2.60E-18 & rs3850634 & 1.90E-15 & rs636497 & rs3850634 & 9 & 91 & 34 & 66 & 51 & 49 \\
   &      &   TC   & 1 &   62870388:63270447   & 4.90E-41 & rs3850634 & 1.90E-15 & rs636497 & rs3850634 & 9 & 90 & 34 & 65 & 51 & 49 \\
   &      &   TG   & 1 &   62870388:63270447   & 8.80E-43 & rs2131925 & 1.90E-15 & rs636497 & rs2131925 & 8 & 92 & 29 & 70 & 45 & 54 \\
DOCK7   & 1.00E-22 &   LDL   & 1 &   62849818:63249877   & 2.60E-18 & rs3850634 & 1.90E-25 & rs11485618 & rs3850634 & 5 & 95 & 23 & 77 & 37 & 63 \\
   &      &      & 1 &   62720869:63120928   & 2.60E-18 & rs3850634 & 0.0049 & rs10458569 & rs3850634 & 1 & 13 & 2 & 3 & 2 & 1 \\
   &      &   TC   & 1 &   62849818:63249877   & 4.90E-41 & rs3850634 & 1.90E-25 & rs11485618 & rs10789118 & 5 & 95 & 20 & 80 & 33 & 67 \\
   &      &      & 1 &   62720869:63120928   & 4.90E-41 & rs3850634 & 0.0049 & rs10458569 & rs3850634 & 1 & 14 & 2 & 3 & 2 & 1 \\
   &      &   TG   & 1 &   62849818:63249877   & 8.80E-43 & rs2131925 & 1.90E-25 & rs11485618 & rs2131925 & 5 & 94 & 23 & 77 & 37 & 63 \\
   &      &      & 1 &   62720869:63120928   & 8.80E-43 & rs2131925 & 0.0049 & rs10458569 & rs2131925 & 1 & 14 & 2 & 3 & 2 & 2 \\
CELSR2   & 5.00E-94 &   LDL   & 1 &   109618271:110018330   & 9.70E-171 & rs629301 & 1.50E-120 & rs646776 & rs629301 &   $<1$   &   $>99$   & 1 & 99 & 2 & 98 \\
   &      &   TC   & 1 &   109618271:110018330   & 5.80E-131 & rs629301 & 1.50E-120 & rs646776 & rs629301 &    $<1$   &   $>99$   & 1 & 99 & 3 & 97 \\
PSMA5   & 9.00E-17 &   LDL   & 1 &   109744528:110144587   & 9.70E-171 & rs629301 & 1.50E-17 & rs599839 & rs629301 &   $<1$   & 99 & 4 & 96 & 7 & 93 \\
   &      &      & 1 &   109741904:110141963   & 9.70E-171 & rs629301 & 1.20E-07 & rs600806 & rs629301 & 98 & 1 & 99 &   $<1$   & 99 &   $<1$   \\
   &      &   TC   & 1 &   109744528:110144587   & 5.80E-131 & rs629301 & 1.50E-17 & rs599839 & rs629301 &   $<1$   & 99 & 3 & 97 & 7 & 93 \\
   &      &      & 1 &   109741904:110141963   & 5.80E-131 & rs629301 & 1.20E-07 & rs600806 & rs629301 & 98 & 1 & 99 &   $<1$   & 99 &   $<1$   \\
PSRC1   & 2.00E-271 &   LDL   & 1 &   109622208:110022267   & 9.70E-171 & rs629301 & 1.10E-299 & rs7528419 & rs629301 &   $<1$   & 99 & 3 & 97 & 7 & 93 \\
   &      &   TC   & 1 &   109622208:110022267   & 5.80E-131 & rs629301 & 1.10E-299 & rs7528419 & rs629301 &   $<1$   & 99 & 3 & 96 & 7 & 93 \\
SORT1   & 2.00E-300 &   LDL   & 1 &   109652373:110052432   & 9.70E-171 & rs629301 & 1.10E-299 & rs7528419 & rs629301 &   $<1$   &  $>99$  & 3 & 96 & 7 & 93 \\
   &      &      & 1 &   109656429:110056488   & 9.70E-171 & rs629301 & 1.10E-299 & rs7528419 & rs629301 &   $<1$   &   $>99$   & 3 & 96 & 7 & 93 \\
   &      &   TC   & 1 &   109652373:110052432   & 5.80E-131 & rs629301 & 1.10E-299 & rs7528419 & rs629301 &   $<1$   &   $>99$   & 3 & 96 & 7 & 93 \\
   &      &      & 1 &   109656429:110056488   & 5.80E-131 & rs629301 & 1.10E-299 & rs7528419 & rs629301 &   $<1$   &   $>99$   & 3 & 96 & 7 & 93 \\
SYPL2   & 1.00E-23 &   LDL   & 1 &   109824678:110224737   & 2.20E-64 & rs672569 & 7.10E-103 & rs1933182 & rs1933182 &   $>99$   &   $<1$   &   $>99$   &   $<1$   &   $>99$   &   $<1$   \\
   &      &      & 1 &   109821999:110222054   & 2.90E-168 & rs599839 & 0.0031 & rs7536292 & rs599839 &   $<1$   &   $<1$   &   $<1$   &   $<1$   &   $<1$   &   $<1$   \\
   &      &   TC   & 1 &   109824678:110224737   & 8.00E-52 & rs672569 & 7.10E-103 & rs1933182 & rs1933182 &   $>99$   &   $<1$   &   $>99$   &   $<1$   &   $>99$   &   $<1$   \\
   &      &      & 1 &   109821999:110222054   & 4.10E-130 & rs599839 & 0.0031 & rs7536292 & rs599839 &   $<1$   &   $<1$   &   $<1$   &   $<1$   &   $<1$   &   $<1$   \\
IFT172   & 7.00E-32 &   TC   & 2 &   27467244:27867303   & 7.30E-27 & rs1260326 & 1.70E-130 & rs704791 & rs704791 &   $>99$   &   $<1$   &   $>99$   &   $<1$   &   $>99$   &   $<1$   \\
   &      &   TG   & 2 &   27467244:27867303   & 5.70E-133 & rs1260326 & 1.70E-130 & rs704791 & rs1260326 &   $>99$   &   $<1$   &   $>99$   &   $<1$   &   $>99$   &   $<1$   \\
SLC39A8   & 3.00E-19 &   HDL   & 4 &   102982958:103383017   & 7.20E-11 & rs13107325 & 3.80E-21 & rs13107325 & rs13107325 &   $<1$   &   $>99$   &   $<1$   &   $>99$   &   $<1$   &   $>99$   \\
   &      &      & 4 &   102972446:103372505   & 7.20E-11 & rs13107325 & 0.027 & rs11733483 & rs13107325 &   $<1$   &   $<1$   &   $<1$   &   $<1$   &   $<1$   &   $<1$   \\
HLA-DQB1   & 2.00E-13 &   TC   & 6 &   32427977:32828028   & 2.60E-17 & rs17533167 & 2.10E-43 & rs3129720 & rs3129720 &   $>99$   &   $<1$   &   $>99$   &   $<1$   &   $>99$   &   $<1$   \\
HLA-DRB1   & 7.00E-44 &   TC   & 6 &   32346553:32746610   & 4.00E-19 & rs3177928 & 1.90E-217 & rs477515 & rs477515 &   $>99$   &   $<1$   &   $>99$   &   $<1$   &   $>99$   &   $<1$   \\
FRK   & 4.00E-12 &   LDL   & 6 &   116062804:116462863   & 2.90E-09 & rs11153594 & 6.60E-15 & rs195517 & rs195517 & 90 & 10 & 98 & 2 & 99 & 1 \\
   &      &   TC   & 6 &   116062804:116462863   & 1.70E-10 & rs9488822 & 6.60E-15 & rs195517 & rs9488822 & 61 & 39 & 89 & 11 & 94 & 6 \\
PPP1R3B   & 1.00E-14 &   LDL   & 8 &   8795514:9195573   & 7.40E-15 & rs2126259 & 6.20E-17 & rs2126259 & rs2126259 &   $<1$   &   $>99$   & 2 & 98 & 4 & 96 \\
   &      &      & 8 &   8793929:9193988   & 7.40E-15 & rs2126259 & 1.90E-17 & rs4240624 & rs9987289 & 1 & 99 & 5 & 94 & 10 & 89 \\
   &      &   TC   & 8 &   8795514:9195573   & 9.00E-24 & rs2126259 & 6.20E-17 & rs2126259 & rs2126259 &   $<1$   &   $>99$   & 2 & 98 & 3 & 97 \\
   &      &      & 8 &   8793929:9193988   & 9.00E-24 & rs2126259 & 1.90E-17 & rs4240624 & rs2126259 & 1 & 98 & 8 & 92 & 14 & 85 \\
   &      &   HDL   & 8 &   8793929:9193988   & 6.40E-25 & rs9987289 & 6.20E-17 & rs2126259 & rs9987289 &   $<1$   &   $>99$   & 2 & 98 & 4 & 96 \\
   &      &      & 8 &   8795514:9195573   & 6.40E-25 & rs9987289 & 1.90E-17 & rs4240624 & rs9987289 & 1 & 98 & 7 & 93 & 14 & 86 \\
TTC39B   & 2.00E-15 &   HDL   & 9 &   14971602:15371661   & 1.30E-13 & rs643531 & 8.10E-18 & rs581080 & rs686030 & 2 & 98 & 10 & 89 & 19 & 81 \\
SPTY2D1   & 1.00E-16 &   TC   & 11 &   18429356:18829415   & 2.50E-08 & rs10832963 & 7.20E-17 & rs10832963 & rs10832963 &    $<1$   & 99 & 3 & 97 & 6 & 94 \\
   &      &      & 11 &   18427988:18828047   & 2.50E-08 & rs10832963 & 2.90E-19 & rs10832963 & rs10832963 &    $<1$   & 99 & 3 & 97 & 6 & 94 \\
FADS1   & 5.00E-18 &   LDL   & 11 &   61367291:61767350   & 1.20E-21 & rs174583 & 2.90E-20 & rs102275 & rs102275 & 2 & 98 & 9 & 91 & 17 & 83 \\
   &      &   TC   & 11 &   61367291:61767350   & 2.10E-22 & rs174550 & 2.90E-20 & rs102275 & rs102275 & 1 & 99 & 7 & 93 & 13 & 87 \\
   &      &   TG   & 11 &   61367291:61767350   & 5.40E-24 & rs174546 & 2.90E-20 & rs102275 & rs102275 & 1 & 99 & 5 & 95 & 10 & 90 \\
   &      &   HDL   & 11 &   61367291:61767350   & 1.50E-22 & rs174601 & 2.90E-20 & rs102275 & rs102275 &   $<1$   & 99 & 4 & 96 & 8 & 92 \\
ST3GAL4   & 2.00E-22 &   LDL   & 11 &   126084467:126484526   & 1.20E-15 & rs11220462 & 7.20E-25 & rs4307732 & rs4307732 & 2 & 98 & 8 & 92 & 15 & 85 \\
   &      &   TC   & 11 &   126084467:126484526   & 2.10E-11 & rs11220463 & 7.20E-25 & rs4307732 & rs7951028 & 1 & 99 & 7 & 93 & 13 & 87 \\
MMAB   & 2.00E-44 &   HDL   & 12 &   109793364:110193423   & 6.90E-15 & rs7134594 & 1.00E-38 & rs7954144 & rs7954144 & 2 & 97 & 12 & 88 & 21 & 79 \\
NYNRIN   & 3.00E-46 &   LDL   & 14 &   24688259:25088318   & 4.40E-11 & rs2332328 & 1.10E-78 & rs6573778 & rs6573778 &   $<1$   & 99 & 3 & 97 & 5 & 95 \\
CKMT1A   & 8.00E-28 &   TG   & 15 &   43791310:44191369   & 4.70E-08 & rs9989313 & 1.20E-13 & rs2260160 & rs9989313 & 5 & 95 & 20 & 79 & 33 & 66 \\
ALDH1A2   & 5.00E-08 &   TC   & 15 &   58334099:58734158   & 8.80E-20 & rs1532085 & 5.00E-45 & rs1532085 & rs1532085 &   $<1$   &   $>99$   & 1 & 99 & 2 & 98 \\
   &      &      & 15 &   58045824:58445883   & 0.0042 & rs4646590 & 0.0027 & rs9796698 & rs9796698 &   $<1$   &   $<1$   &   $<1$   &   $<1$   &   $<1$   &   $<1$   \\
   &      &   HDL   & 15 &   58334099:58734158   & 2.90E-96 & rs1532085 & 5.00E-45 & rs1532085 & rs1532085 &   $<1$   &   $>99$   &   $<1$   &   $>99$   & 1 & 99 \\
   &      &      & 15 &   58045824:58445883   & 0.0014 & rs6493966 & 0.0027 & rs9796698 & rs8040266 &   $<1$   &   $<1$   &   $<1$   &   $<1$   &   $<1$   &   $<1$   \\
   &      &   TG   & 15 &   58334099:58734158   & 2.40E-13 & rs261342 & 5.00E-45 & rs1532085 & rs2043085 & 4 & 96 & 18 & 81 & 31 & 69 \\
   &      &      & 15 &   58045824:58445883   & 0.0046 & rs2414536 & 0.0027 & rs9796698 & rs9796698 &   $<1$   &   $<1$   &   $<1$   &   $<1$   &   $<1$   &   $<1$   \\
LIPC   & 7.00E-23 &   TC   & 15 &   58653103:59053162   & 8.80E-20 & rs1532085 & 1.10E-25 & rs2043085 & rs2043085 &   $<1$   &   $>99$   & 1 & 99 & 3 & 97 \\
   &      &   HDL   & 15 &   58653103:59053162   & 2.90E-96 & rs1532085 & 1.10E-25 & rs2043085 & rs1532085 &   $<1$   &   $>99$   & 2 & 98 & 3 & 97 \\
   &      &   TG   & 15 &   58653103:59053162   & 2.40E-13 & rs261342 & 1.10E-25 & rs2043085 & rs2043085 & 2 & 98 & 10 & 89 & 19 & 81 \\
VKORC1   & 7.00E-47 &   TG   & 16 &   30904631:31304682   & 3.30E-08 & rs11649653 & 1.10E-80 & rs749671 & rs749671 & 3 & 97 & 13 & 87 & 22 & 77 \\
NFATC3   & 3.00E-15 &   HDL   & 16 &   68060252:68460297   & 1.20E-27 & rs3785108 & 3.30E-05 & rs7204208 & rs3785108 & 3 & 91 & 10 & 68 & 15 & 51 \\
   &      &      & 16 &   68051875:68451934   & 1.20E-27 & rs3785108 & 0.0067 & rs3743739 & rs3785108 &   $<1$   & 5 &   $<1$   &   $<1$   &   $<1$   &   $<1$   \\
   &      &      & 16 &   68025772:68425831   & 1.20E-27 & rs3785108 & 0.00055 & rs9930867 & rs3785108 & 2 &   $<1$   & 2 &   $<1$   & 2 &   $<1$   \\
   &      &      & 16 &   68063016:68463075   & 1.20E-27 & rs3785108 & 0.0068 & rs1868158 & rs3785108 &   $<1$   &   $<1$   &   $<1$   &   $<1$    &   $<1$   &   $<1$   \\
   &      &      & 16 &   68060703:68460762   & 1.20E-27 & rs3785108 & 4.20E-30 & rs7193701 & rs7204208 &   $>99$   &   $<1$   &   $>99$   &   $<1$   &   $>99$   &   $<1$   \\
PERLD1   & 9.00E-24 &   HDL   &   NA   &   NA   &   NA   &   NA   &   NA   &   NA   &   NA   &   NA   &   NA   &   NA   &   NA   &   NA   &   NA   \\
TBKBP1   & 6.00E-10 &   TC   & 17 &   45589357:45989416   & 1.80E-07 & rs8072100 & 2.10E-21 & rs9913503 & rs4794053 & 78 & 17 & 90 & 4 & 92 & 2 \\
   &      &   LDL   & 17 &   45589357:45989416   & 1.10E-07 & rs8072100 & 2.10E-21 & rs9913503 & rs6503807 & 47 & 51 & 79 & 17 & 87 & 9 \\
LIPG   & 4.00E-10 &   TC   & 18 &   46918514:47318573   & 2.00E-19 & rs7239867 & 1.20E-11 & rs4939883 & rs4939883 & 1 & 99 & 5 & 94 & 11 & 89 \\
   &      &   HDL   & 18 &   46918514:47318573   & 2.70E-49 & rs7241918 & 1.20E-11 & rs4939883 & rs4939883 &    $<1$   &   $>99$   & 2 & 98 & 4 & 96 \\
ANGPTL4   & 4.00E-08 &   HDL   & 19 &   8239194:8639253   & 3.20E-08 & rs7255436 & 3.80E-09 & rs7255436 & rs7255436 &   $<1$   &   $>99$   & 2 & 98 & 5 & 95 \\
APOC4   & 4.00E-09 &   TG   & 19 &   45248464:45648523   & 1.10E-30 & rs439401 & 9.40E-15 & 19:45430280 & rs439401 & 98 & 2 &   $>99$   &   $<1$   &   $>99$   &   $<1$   \\
   &      &      & 19 &   45252653:45652712   & 1.10E-30 & rs439401 & 1.10E-299 & rs1130742 & rs1130742 &   $>99$   &   $<1$   &   $>99$   &   $<1$   &   $>99$   &   $<1$   \\
LILRA3   & 9.00E-12 &   HDL   & 19 &   54602026:55002085   & 4.30E-16 & rs386000 & 8.20E-17 & 19:54793830 & rs386000 &   $<1$   &   $>99$   & 4 & 95 & 9 & 91 \\
CEP250   & 3.00E-08 &   TC   & 20 &   33899702:34299761   & 3.80E-10 & rs2277862 & 0.022 & rs17424259 & rs2104417 &   $<1$   &   $<1$   &   $<1$   &   $<1$   &   $<1$   &   $<1$   \\
CPNE1   & 7.00E-41 &   TC   & 20 &   34013995:34414054   & 3.80E-10 & rs2277862 & 7.30E-110 & rs6060524 & rs6060524 & 95 & 5 & 99 & 1 & 99 &   $<1$   \\
PLTP   & 3.00E-18 &   TG   & 20 &   44327404:44727463   & 4.70E-18 & rs4810479 & 1.80E-20 & rs6065906 & rs4810479 &   $<1$   &   $>99$   & 3 & 97 & 5 & 95 \\
   &      &   HDL   & 20 &   44327404:44727463   & 1.90E-22 & rs6065906 & 1.80E-20 & rs6065906 & rs6065906 &   $<1$   &   $>99$   & 2 & 98 & 5 & 95 \\
UBE2L3   & 6.00E-13 &   HDL   & 22 &   21778264:22178323   & 1.10E-08 & 22:21932068 & 8.90E-13 & rs4821112 & rs2283790 & 3 & 96 & 16 & 84 & 27 & 72 \\ [1ex] 
   \hline  
\end{longtable}
\begin{flushleft}Published results of loci correlating with both liver expression and one of the four lipid traits (Teslovich et al. Supplementary Table 8) and posterior probability of different signal (PP3) and common signal (PP4) after applying colocalisation test. Each row lists the results for one probe, and the multiple entries for the same locus and trait represent multiple probes mapping to the same locus. the columns \textbf{Biom pval} and \textbf{eQTL pval} report the lowest p-value found for the association with the trait listed and for the liver expression association respectively, with the corresponding SNP name (\textbf{Biom SNP} and \textbf{eQTL SNP}); the column \textbf{Best Causal} reports the SNP within the region with the highest posterior probability to be the true causal variant. The probabilities have been rounded to 1 significant figure. 
\end{flushleft}
    \label{tesl_table}
 \end{landscape}
\clearpage

\newpage
\begin{table}[!ht]   
\caption{
\bf{eQTL/LDL colocalisation}}
\resizebox{\columnwidth}{!}{%
\begin{tabular}{lllllllllll}
\tabularnewline \hline
\hline 
Chr  &  Region  &  Signal  &  PP.H3.abf  &  PP.H4.abf  &  Tesl   &  Biom pval  &  Biom SNP   &  eQTL pval  &  eQTL SNP  &  Best Causal  \\
\hline
1 &   109618271:110144587   &  CELSR2  & 2 & 98 &   Y   & 9.70E-171 &   rs629301   & 1.50E-120 &   rs646776   &   rs629301   \\
  &    &  PSRC1  & 7 & 93 &   Y   & 9.70E-171 &   rs629301   & 1.10E-299 &   rs7528419   &   rs629301   \\
  &    &  SORT1  & 7 & 93 &   Y   & 9.70E-171 &   rs629301   & 1.10E-299 &   rs7528419   &   rs629301   \\
  &    &  PSMA5  & 7 & 93 &   Y   & 9.70E-171 &   rs629301   & 1.50E-17 &   rs599839   &   rs629301   \\
1 &  150767063:151167122  &  ANXA9  & 3 & 97 &   N   & 6.50E-08 &   rs267733   & 6.10E-08 &   rs267734   &   rs267733   \\
1 &  25626305:26026364  &  TMEM57  & 10 & 90 &   Y   & 1.20E-10 &   rs12027135   & 2.10E-31 &   rs873308   &   rs12027135   \\
2 &  120908798:121308857  &  INHBB  & 7 & 77 &   N   & 3.80E-07 &   rs2030746   & 4.90E-21 &   rs17050272   &   rs17050272   \\
8 &  59158506:59558565  &  UBXN2B  & 13 & 87 &   N   & 3.90E-09 &   rs1030431   & 3.50E-10 &   rs11996829   &   rs13263105   \\
8 &   8795514:9392282   &  PPP1R3B  & 4 & 96 &   Y   & 7.40E-15 &   rs2126259   & 6.20E-17 &   rs2126259   &   rs2126259   \\
  &    &  ENSG00000254235  & 13 & 87 &   N   & 7.40E-15 &   rs2126259   & 4.00E-13 &   rs4841133   &   rs9987289   \\
9 &  2454062:2854121  &  VLDLR  & 1 & 91 &   N   & 8.00E-06 &   rs3780181   & 1.40E-07 &   rs3780181   &   rs3780181   \\
11 &  126084467:126484526  &  ST3GAL4  & 15 & 85 &   Y   & 1.20E-15 &   rs11220462   & 7.20E-25 &   rs4307732   &   rs4307732   \\
11 &  18429356:18829415  &  SPTY2D1  & 6 & 93 &   N   & 2.90E-07 &   rs10128711   & 7.20E-17 &   rs10832963   &   rs10832963   \\
11 &  61367291:61767350  &  FADS1  & 17 & 83 &   Y   & 1.20E-21 &   rs174583   & 2.90E-20 &   rs102275   &   rs102275   \\
12 &  111508189:111908248  &  CUX2  & 2 & 98 &   N   & 1.50E-09 &   rs11065987   & 1.00E-17 &   rs4378452   &   rs3184504   \\
14 &  24688259:25088318  &  NYNRIN  & 5 & 95 &   Y   & 4.40E-11 &   rs2332328   & 1.10E-78 &   rs6573778   &   rs6573778   \\
16 &   71894416:72310900   &  HP  & 1 & 97 &   N   & 1.80E-22 &   rs2000999   & 1.30E-05 &   rs2000999   &   rs2000999   \\
  &    &  HPR  & 1 & 99 &   N   & 1.80E-22 &   rs2000999   & 4.20E-08 &   rs2000999   &   rs2000999   \\
17 &  45562645:45962704  &  KPNB1  & 16 & 83 &   N   & 1.10E-07 &   rs8072100   & 3.10E-09 &   rs4794048   &   rs8072100   \\
20 &  12947054:13347113  &  SPTLC3  & 4 & 86 &   N   & 4.20E-06 &   rs364585   & 2.30E-41 &   rs168622   &   rs364585  \\
     \hline 
    \end{tabular} 
    }
  \begin{flushleft} Positive (PP4 $> 75\%$) eQTL/LDL colocalisation results between the liver eQTL dataset and the Teslovich meta-analysis using the most stringent prior for the probability that one SNP is associated with both traits, $p_{12}=10^{-6}$. The column \textbf{Signal} includes genes that are part of overlapping regions and that colocalise at  PP4 $> 75\%$; the column \textbf{Region} represents the genomic coordinates for the start and stop of the signal; in the column \textbf{Tesl}, ``Y" indicates that this signal with any of the genes included has been reported to be an intermediate for any of the four lipid biomarker associations by Teslovich et al. ; the columns \textbf{Biom pval} and \textbf{eQTL pval} report the lowest p-value found for LDL association and for the expression association respectively, with the corresponding SNP name (\textbf{Biom SNP} and \textbf{eQTL SNP}); the column \textbf{Best Causal} reports the SNP within the region with the highest posterior probability to be the true causal variant. The probabilities have been rounded to 1 significant figure. 
\end{flushleft}
  \label{best.ldl}
\end{table}
\clearpage

\newpage
\begin{table}[!ht]   
\caption{
\bf{eQTL/HDL colocalisation}}
\resizebox{\columnwidth}{!}{%
\begin{tabular}{lllllllllll}
\tabularnewline \hline
\hline 
Chr  &  Region  &  Signal  &  PP.H3.abf  &  PP.H4.abf  &  Tesl   &  Biom pval  &  Biom SNP   &  eQTL pval  &  eQTL SNP  &  Best Causal  \\
\hline
1 &  109618271:110144587  &  CELSR2  & 3 & 97 &  N  & 6.20E-08 &  rs629301  & 1.50E-120 &  rs646776  &  rs646776  \\
  &    &  PSRC1  & 7 & 93 &  N  & 6.20E-08 &  rs629301  & 1.10E-299 &  rs7528419  &  rs629301  \\
  &    &  SORT1  & 7 & 93 &  N  & 6.20E-08 &  rs629301  & 1.10E-299 &  rs7528419  &  rs629301  \\
  &    &  PSMA5  & 7 & 92 &  N  & 6.20E-08 &  rs629301  & 1.50E-17 &  rs599839  &  rs12740374  \\
2 &  85349026:85749085  &  TGOLN2  & 17 & 83 &   N   & 1.00E-07 &   rs1053560   & 2.80E-80 &   rs1044973   &   rs1044973   \\
4 &  102982958:103383017  &  SLC39A8  & 1 & 99 &   Y   & 7.20E-11 &   rs13107325   & 3.80E-21 &   rs13107325   &   rs13107325   \\
8 &   8795514:9392282   &  PPP1R3B  & 14 & 86 &   Y   & 6.40E-25 &   rs9987289   & 1.90E-17 &   rs4240624   &   rs9987289   \\
  &    &  ENSG00000254235  & 4 & 96 &   N   & 6.40E-25 &   rs9987289   & 4.00E-13 &   rs4841133   &   rs9987289   \\
9 &  14971602:15371661  &  TTC39B  & 19 & 81 &   Y   & 1.30E-13 &   rs643531   & 8.10E-18 &   rs581080   &   rs686030   \\
11 &  61367291:61767350  &  FADS1  & 8 & 92 &   Y   & 1.50E-22 &   rs174601   & 2.90E-20 &   rs102275   &   rs102275   \\
12 &  109793364:110193423  &  MMAB  & 21 & 79 &   Y   & 6.90E-15 &   rs7134594   & 1.00E-38 &   rs7954144   &   rs7954144   \\
12 &  111508189:111908248  &  CUX2  & 2 & 89 &   N   & 4.40E-06 &   rs4766578   & 2.80E-16 &   rs3184504   &   rs3184504   \\
15 &   58334099:59053162   &  ALDH1A2  & 1 & 99 &   Y   & 2.90E-96 &   rs1532085   & 5.00E-45 &   rs1532085   &   rs1532085   \\
  &    &  LIPC  & 3 & 97 &   Y   & 2.90E-96 &   rs1532085   & 1.10E-25 &   rs2043085   &   rs1532085   \\
15 &  96517293:96917352  &  ENSG00000259359  & 2 & 87 &   N   & 8.00E-06 &   rs8023580   & 5.50E-13 &   rs8023580   &   rs8023580   \\
18 &  46918514:47318573  &  LIPG  & 4 & 96 &   Y   & 2.70E-49 &   rs7241918   & 1.20E-11 &   rs4939883   &   rs4939883   \\
19 &   54578320:55002085   &  LILRB2  & 9 & 88 &   N   & 4.30E-16 &   rs386000   & 1.70E-06 &   rs416867   &   rs386000   \\
  &    &  LILRA3  & 9 & 91 &   Y   & 4.30E-16 &   rs386000   & 8.20E-17 &   19:54793830   &   rs386000   \\
19 &  8239194:8639253  &  ANGPTL4  & 5 & 95 &   Y   & 3.20E-08 &   rs7255436   & 3.80E-09 &   rs7255436   &   rs7255436   \\
20 &  44327404:44727463  &  PLTP  & 5 & 95 &   Y   & 1.90E-22 &   rs6065906   & 1.80E-20 &   rs6065906   &   rs6065906 \\
     \hline 
    \end{tabular} 
    }
  \begin{flushleft} Positive (PP4 $> 75\%$) eQTL/HDL colocalisation results between the liver eQTL dataset and the Teslovich meta-analysis. Column and row headings are the same as in previous figure. 
\end{flushleft}
  \label{best.hdl}
\end{table}
\clearpage

\newpage
\begin{table}[!ht]   
\caption{
\bf{eQTL/TG colocalisation}}
\resizebox{\columnwidth}{!}{%
\begin{tabular}{lllllllllll}
\tabularnewline \hline
\hline 
Chr  &  Region  &  Signal  &  PP.H3.abf  &  PP.H4.abf  &  Tesl   &  Biom pval  &  Biom SNP   &  eQTL pval  &  eQTL SNP  &  Best Causal  \\
\hline
2 &   2754647:28005583   &  GCKR  & 5 & 77 &   N   & 5.70E-133 &   rs1260326   & 1.50E-05 &   rs1260326   &   rs1260326   \\
  &    &  C2orf16  & 4 & 81 &   N   & 5.70E-133 &   rs1260326   & 8.30E-06 &   rs1260326   &   rs1260326   \\
10 &  94637063:95037122  &  CYP26A1  & 3 & 95 &   N   & 2.40E-08 &   rs2068888   & 3.50E-06 &   rs4418728   &   rs2068888   \\
11 &  61367291:61767350  &  FADS1  & 10 & 90 &   Y   & 5.40E-24 &   rs174546   & 2.90E-20 &   rs102275   &   rs102275   \\
15 &  58653103:59053162  &  LIPC  & 19 & 81 &   Y   & 2.40E-13 &   rs261342   & 1.10E-25 &   rs2043085   &   rs2043085   \\
16 &  30904631:31304682  &  VKORC1  & 23 & 77 &   Y   & 3.30E-08 &   rs11649653   & 1.10E-80 &   rs749671   &   rs749671   \\
  &    &   ENSG00000255439   & 23 & 77 &   N   & 3.30E-08 &   rs11649653   & 1.10E-80 &   rs749671   &   rs749671   \\
16 &  71894416:72310900  &  HP  & 2 & 75 &   N   & 5.70E-06 &   rs2000999   & 2.10E-06 &   rs2000999   &   rs2000999   \\
  &    &  HPR  & 2 & 89 &   N   & 5.70E-06 &   rs2000999   & 4.20E-08 &   rs2000999   &   rs2000999   \\
20 &  44327404:44727463  &  PLTP  & 5 & 95 &   Y   & 4.70E-18 &   rs4810479   & 1.80E-20 &   rs6065906   &   rs4810479 \\
     \hline 
    \end{tabular} 
    }
  \begin{flushleft} Positive (PP4 $> 75\%$) eQTL/TG colocalisation results between the liver eQTL dataset and the Teslovich meta-analysis. Column and row headings are the same as in previous figure.
\end{flushleft}
  \label{best.tg}
\end{table}
\clearpage

\newpage
\begin{table}[!ht]   
\caption{
\bf{eQTL/TC colocalisation}}
\resizebox{\columnwidth}{!}{%
\begin{tabular}{lllllllllll}
\tabularnewline \hline
\hline 
Chr  &  Region  &  Signal  &  PP.H3.abf  &  PP.H4.abf  &  Tesl   &  Biom pval  &  Biom SNP   &  eQTL pval  &  eQTL SNP  &  Best Causal  \\
\hline
1 &  109618271:110144587  &  CELSR2  & 3 & 97 &   Y   & 5.80E-131 &   rs629301   & 1.50E-120 &   rs646776   &   rs629301   \\
  &    &  PSRC1  & 7 & 93 &   Y   & 5.80E-131 &   rs629301   & 1.10E-299 &   rs7528419   &   rs629301   \\
  &    &  SORT1  & 7 & 93 &   Y   & 5.80E-131 &   rs629301   & 1.10E-299 &   rs7528419   &   rs629301   \\
  &    &  PSMA5  & 7 & 93 &   Y   & 5.80E-131 &   rs629301   & 1.50E-17 &   rs599839   &   rs629301   \\
1 &  25626305:26026364  &  TMEM57  & 11 & 89 &   Y   & 4.10E-11 &   rs12027135   & 2.10E-31 &   rs873308   &   rs12027135   \\
2 &  20201795:20601854  &  SDC1  & 17 & 82 &   N   & 1.20E-07 &   rs1473886   & 6.70E-09 &   rs907866   &   rs1107851   \\
2 &  27546474:28005583  &  GCKR  & 5 & 77 &   N   & 7.30E-27 &   rs1260326   & 1.50E-05 &   rs1260326   &   rs1260326   \\
  &    &  C2orf16  & 4 & 81 &   N   & 7.30E-27 &   rs1260326   & 8.30E-06 &   rs1260326   &   rs1260326   \\
3 &  32322873:32722932  &  CMTM6  & 8 & 77 &   N   & 9.10E-07 &   rs7640978   & 2.70E-07 &   rs17029597   &   rs17029597   \\
6 &  34355095:34755154  &  C6orf106  & 15 & 85 &   N   & 4.70E-11 &   rs2814982   & 4.50E-09 &   rs3800461   &   rs3800461   \\
8 &  59158506:59558565  &  UBXN2B  & 15 & 85 &   N   & 8.80E-13 &   rs1030431   & 3.50E-10 &   rs11996829   &   rs13263105   \\
8 &  8795514:9195573  &  PPP1R3B  & 3 & 97 &   Y   & 9.00E-24 &   rs2126259   & 6.20E-17 &   rs2126259   &   rs2126259   \\
9 &  14971602:15371661  &  TTC39B  & 1 & 99 &   N   & 3.10E-09 &   rs581080   & 8.10E-18 &   rs581080   &   rs581080   \\
10 &  17079389:17479448  &  VIM  & 5 & 93 &   N   & 7.20E-07 &   rs7903259   & 9.80E-09 &   rs10904908   &   rs7903259   \\
11 &  126084467:126484526  &  ST3GAL4  & 13 & 87 &   Y   & 2.10E-11 &   rs11220463   & 7.20E-25 &   rs4307732   &   rs7951028   \\
11 &  18429356:18829415  &  SPTY2D1  & 6 & 94 &   Y   & 2.50E-08 &   rs10832963   & 7.20E-17 &   rs10832963   &   rs10832963   \\
11 &  61367291:61767350  &  FADS1  & 13 & 87 &   Y   & 2.10E-22 &   rs174550   & 2.90E-20 &   rs102275   &   rs102275   \\
12 &  111508189:111908248  &  CUX2  & 2 & 98 &   N   & 2.40E-11 &   rs4766578   & 2.80E-16 &   rs3184504   &   rs3184504   \\
14 &  24688259:25088318  &  NYNRIN  & 3 & 97 &   N   & 1.10E-07 &   rs6573778   & 1.10E-78 &   rs6573778   &   rs6573778   \\
15 &  58334099:59053162  &  ALDH1A2  & 2 & 98 &   Y   & 8.80E-20 &   rs1532085   & 5.00E-45 &   rs1532085   &   rs1532085   \\
  &    &  LIPC  & 3 & 97 &   Y   & 8.80E-20 &   rs1532085   & 1.10E-25 &   rs2043085   &   rs2043085   \\
16 &  56310220:56710279  &  OGFOD1  & 7 & 84 &    & 3.20E-06 &  rs11644679  & 3.40E-11 &  rs11649379  &  rs11644679  \\
16 &  71894416:72310900  &  HP  & 1 & 97 &   N   & 3.20E-24 &   rs2000999   & 2.10E-06 &   rs2000999   &   rs2000999   \\
  &    &  HPR  & 1 & 99 &   N   & 3.20E-24 &   rs2000999   & 4.20E-08 &   rs2000999   &   rs2000999   \\
17 &  45562645:45962704  &  KPNB1  & 13 & 86 &   N   & 1.80E-07 &   rs8072100   & 3.10E-09 &   rs4794048   &   rs8072100   \\
18 &  46918514:47318573  &  LIPG  & 11 & 89 &   Y   & 2.00E-19 &   rs7239867   & 1.20E-11 &   rs4939883   &   rs4939883   \\
22 &  46433083:46833138  &  PPARA  & 10 & 81 &   N   & 3.60E-06 &   rs4253772   & 6.00E-08 &   rs11704979   &   rs4253772  \\
     \hline 
    \end{tabular} 
    }
  \begin{flushleft} Positive (PP4 $> 75\%$) eQTL/TC colocalisation results between the liver eQTL dataset and the Teslovich meta-analysis. TColumn and row headings are the same as in previous figure. 
\end{flushleft}
  \label{best.tc}
\end{table}
\clearpage

\newgeometry{top = 1.5cm, bottom=0.1cm}
\begin{figure}[htb]  
\centering
      \includegraphics[trim={5.5cm 2cm 6cm 4.5cm},clip]{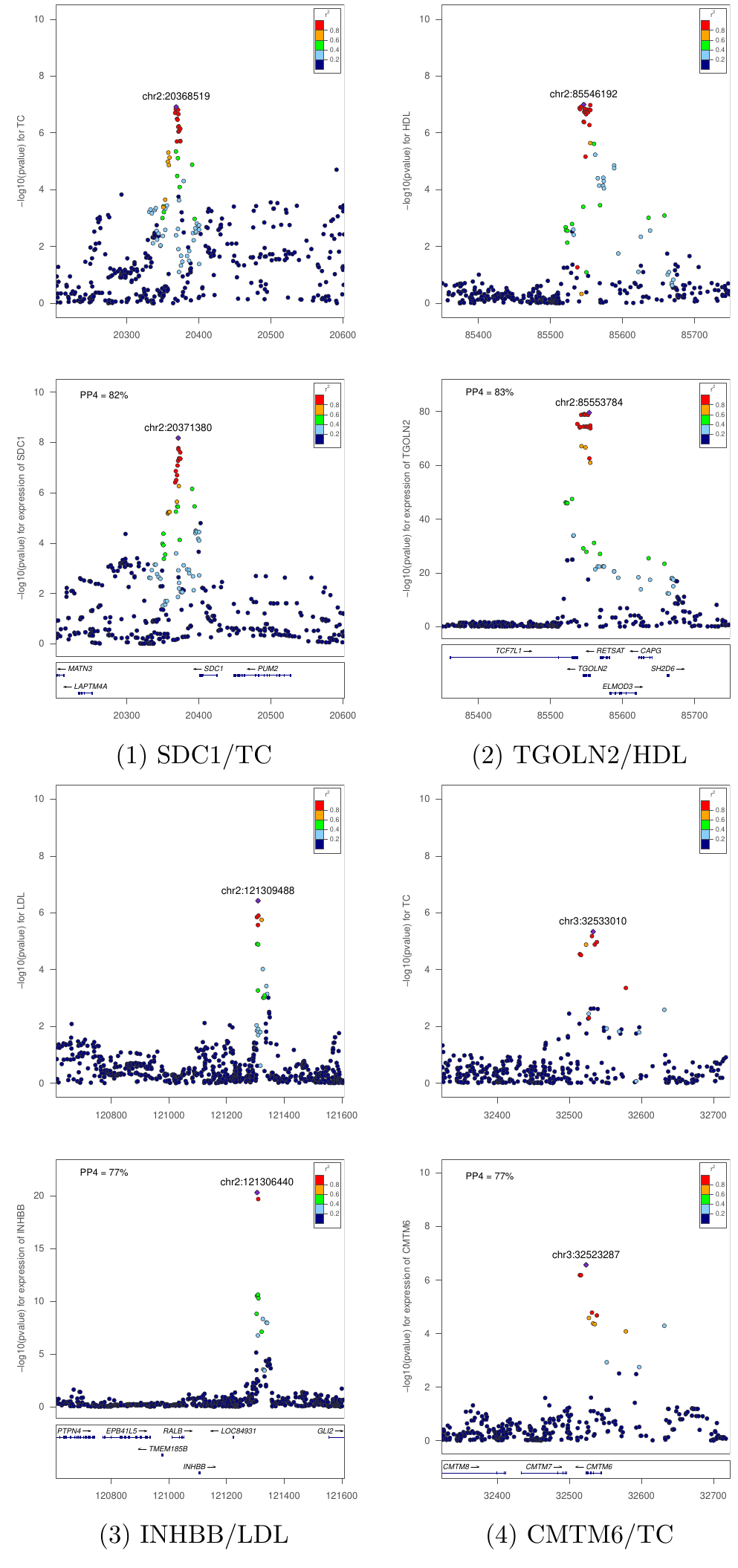} \\ 
  \caption*{}
\end{figure}
\clearpage

\begin{figure}[htb]  
\centering
      \includegraphics[trim={5.5cm 2cm 6cm 4.5cm},clip]{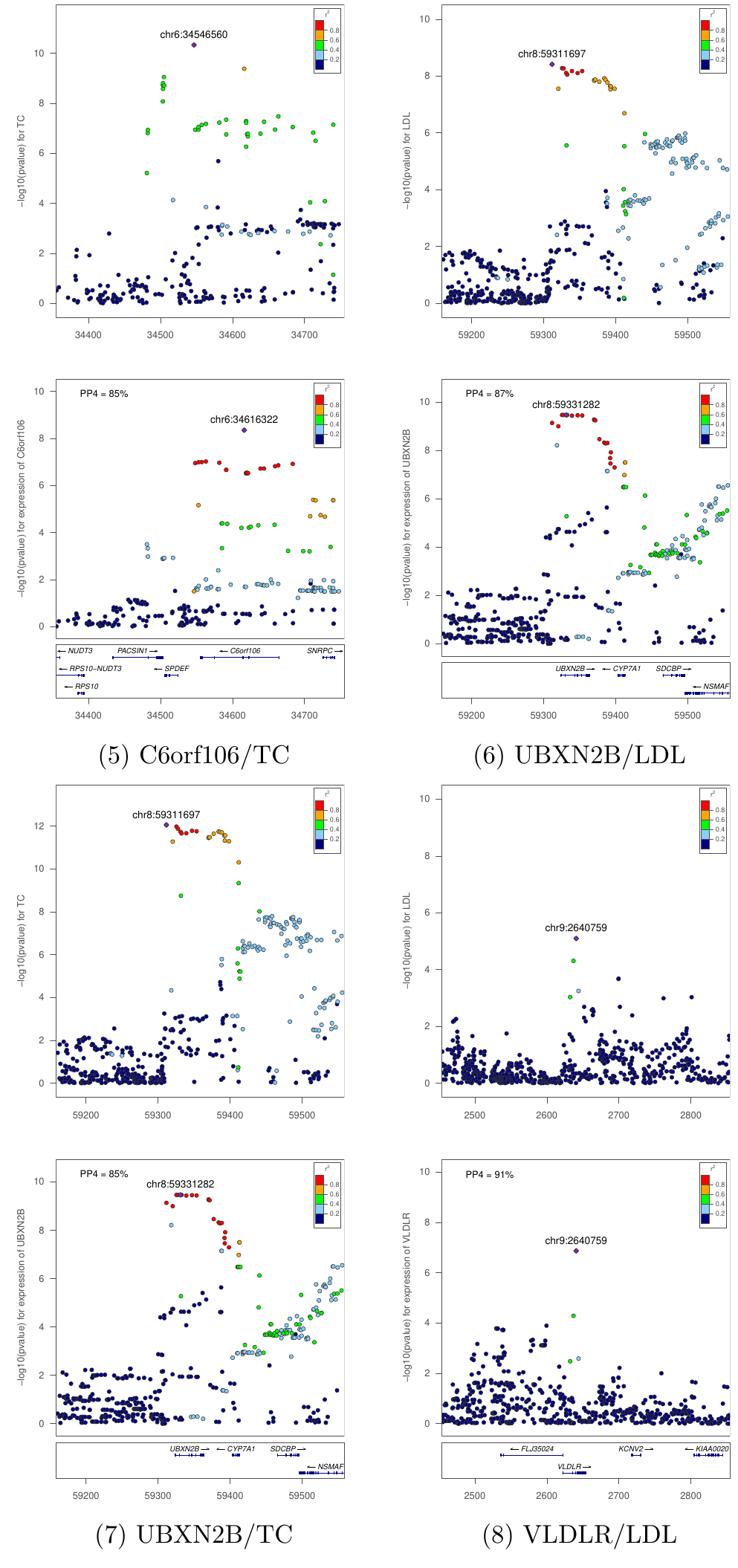} \\ 
  \caption*{}
\end{figure}
\clearpage

\begin{figure}[htb]  
\centering
      \includegraphics[trim={5.5cm 2cm 6cm 4.5cm},clip]{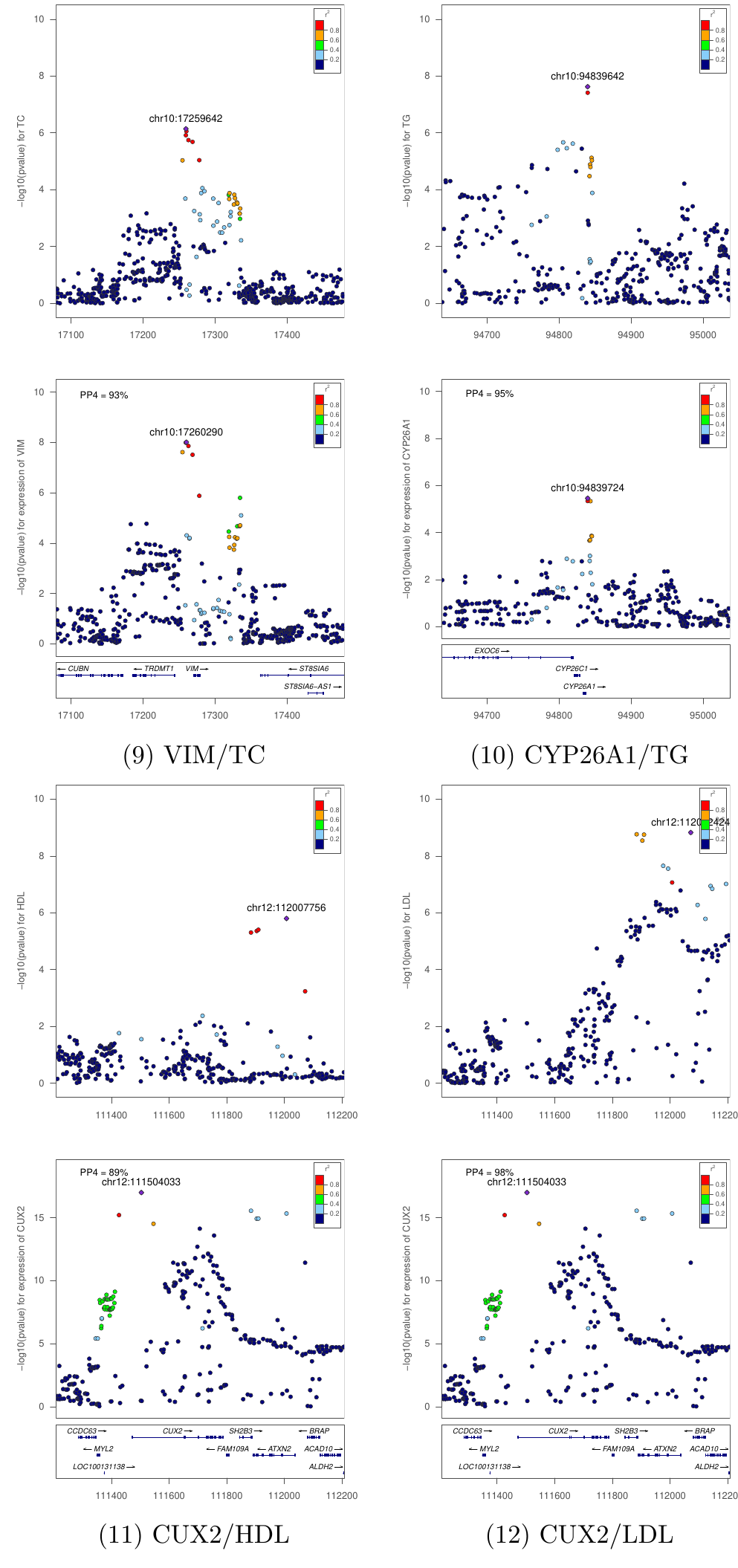} \\ 
  \caption*{}
\end{figure}
\clearpage

\begin{figure}[htb]  
\centering
      \includegraphics[trim={5.5cm 2cm 6cm 4.5cm},clip]{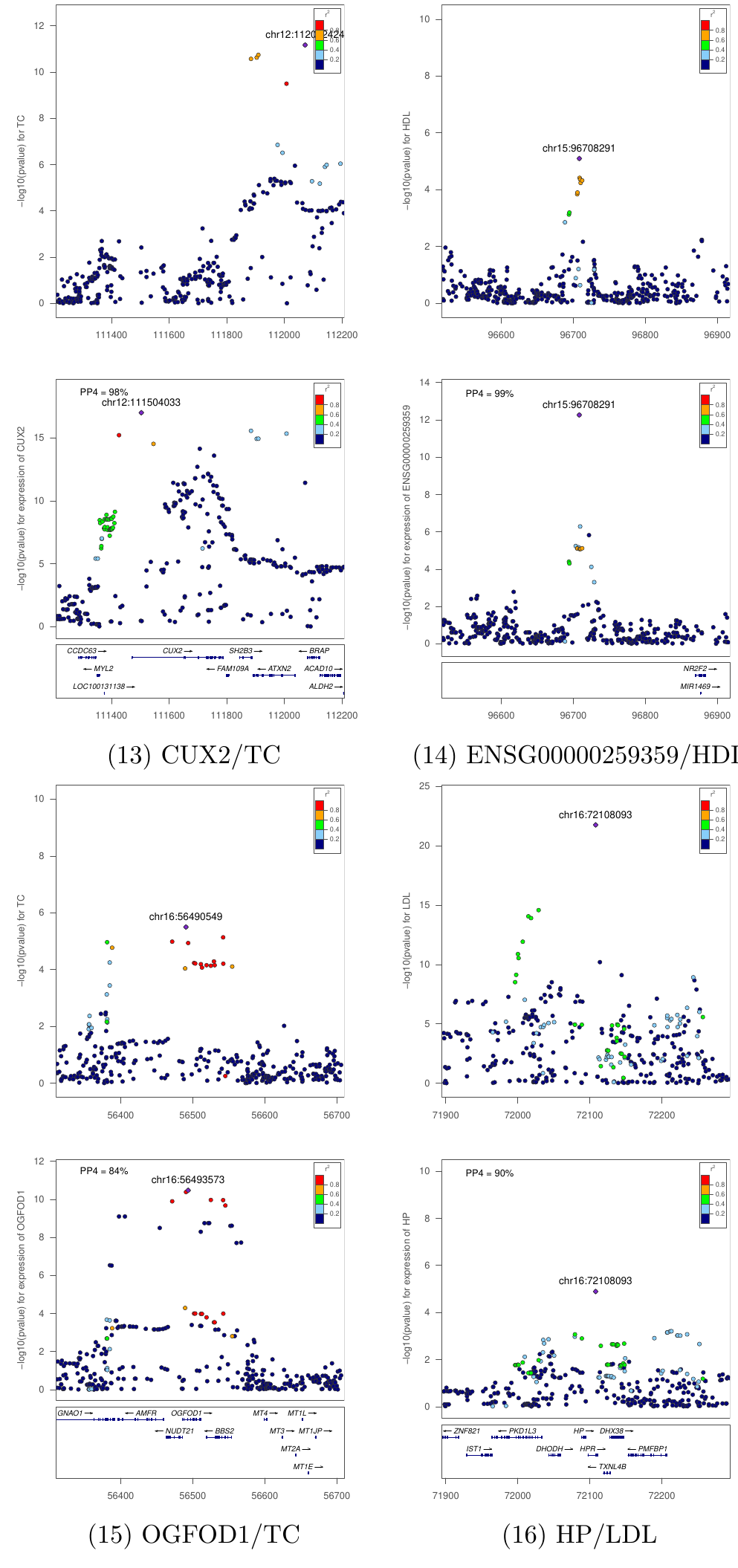} \\ 
  \caption*{}
\end{figure}
\clearpage

\begin{figure}[htb]  
\centering
      \includegraphics[trim={5.5cm 2cm 6cm 4.5cm},clip]{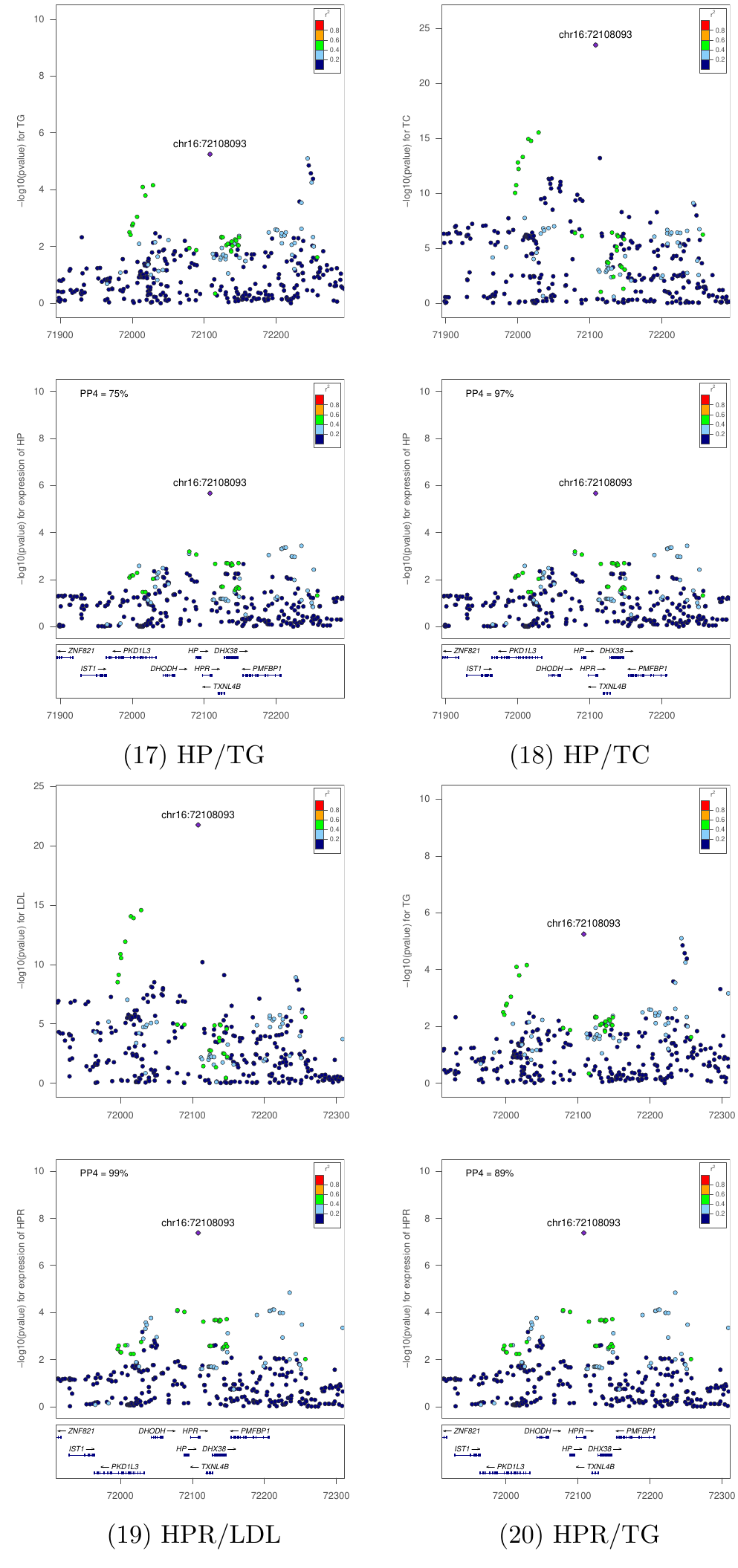} \\ 
  \caption*{}
\end{figure}
\clearpage

\begin{figure}[htb]  
\centering
      \includegraphics[trim={5.5cm 12.5cm 6cm 4cm},clip]{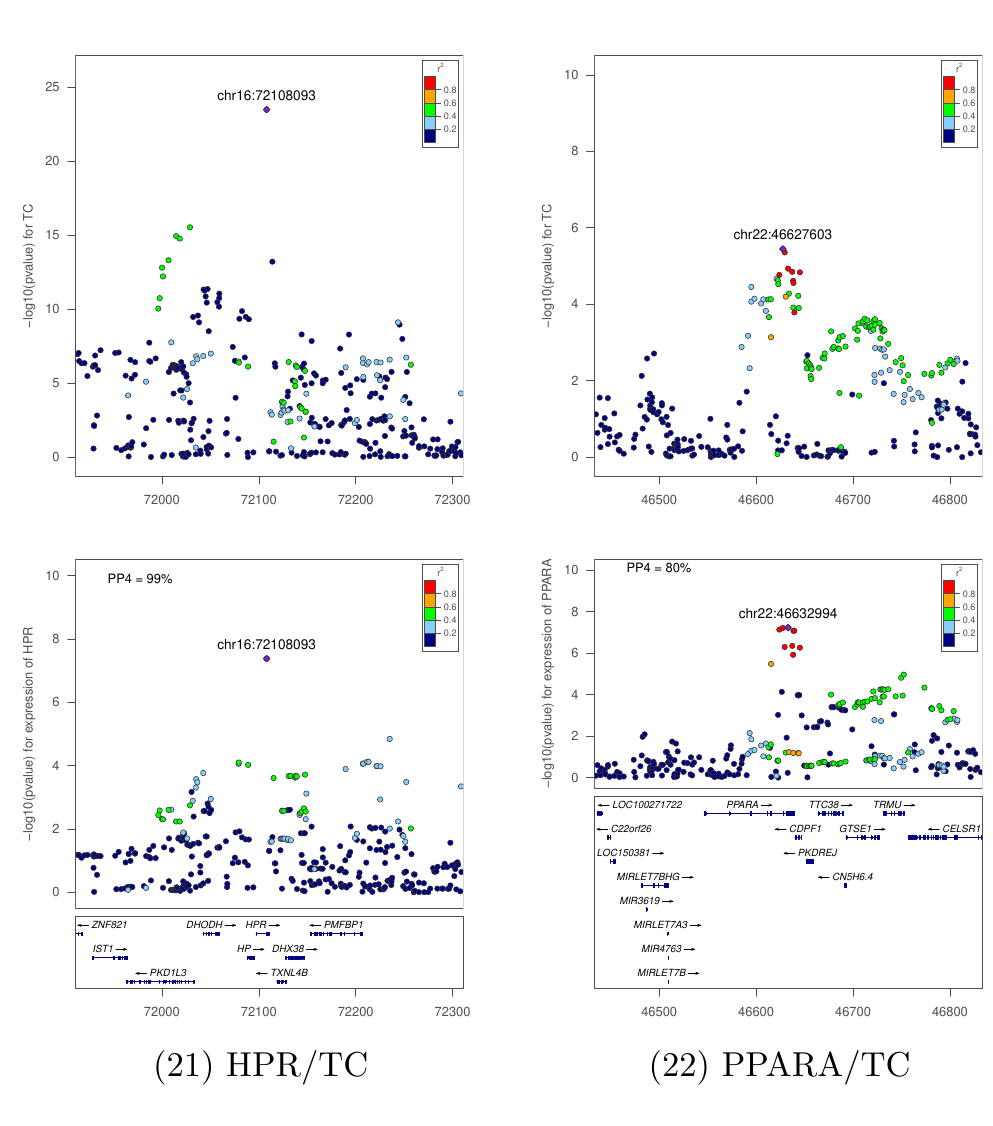} \\ 
       \caption{
 {\bf  Regional Manhattan plots corresponding to loci listed in Table 2 of main text.} Row and column headers defined as in previous figure. The genomic range may be greater than $\sim 400$ kilobases to improve visualisation of the signal.}
  \label{novel_plots}
 \end{figure}
\restoregeometry
\clearpage

\subsection*{Overview of gene function of new colocalisation results associated with blood lipid levels and liver expression}

\begin{itemize}
  \item {\it SDC1} (Syndecan-1) encodes a transmembrane heparan sulfate proteoglycan which mediates the clearance of triglyceride-rich lipoproteins in liver \cite{Yilmaz2012}.

  \item {\it TGOLN2} (Trans-Golgi Network Protein 2) encoded by this gene regulates cholesterol transport to the trans-Golgi network and plasma membrane caveolae \cite{Garver2002}. 
  
  \item {\it INHBB}  encodes Inhibin beta B, a subunit of both activin and inhibin. This locus has been reported associated with elevated levels of lipids and cardiovascular disease risk traits \cite{Johnson2012}.

  \item {\it UBXN2B} (UBX Domain Protein 2B) encodes a protein containing a UBX-domain involved in endoplasmic reticulum-associated degradation, a process which is crucial for lipid droplets maintenance  \cite{Wang2012}. 

  \item {\it VLDLR} (Very Low Density Lipoprotein Receptor) is known to play important roles in VLDL-triglyceride metabolism, and has been previously correlated with glucose and triglyceride plasma levels \cite{Nasarre2012}. 

  \item {\it VIM} codes for Vimentin, and a functional role for vimentin intermediate filaments has been reported in the metabolism of lipoprotein-derived cholesterol \cite{Sarria1992}.

  \item {\it CYP26A1} encodes an endoplasmic reticulum protein which regulates the the cellular level of retinoic acid, a critical signalling molecule involved in the regulation of gene expression. This protein belongs to the cytochrome P450 superfamily of enzymes which catalyse many reactions involved in  maintenance of lipid homeostasis and drug metabolism \cite{Hafner2011}. 

  \item {\it OGFOD1} (2-oxoglutarate and iron-dependent oxygenase domain containing 1) is crucial for cellular adaptation to changes in oxygen concentration, and has been reported to function in ischaemic signalling \cite{Saito2010}.

 \item {\it HP} encodes the plasma protein haptoglobin, which binds and transports free hemoglobin (Hb) released from erythrocytes back to the liver for recycling, thereby inhibiting haemoglobin's oxidative activity \cite{Wassell1999}.
 
 \item Haptoglobin-related protein ({\it HPR}) is a plasma protein associated with apolipoprotein-L-I containing high-density lipoprotein (HDL) particles, and has been shown to be part of the innate immune response \cite{Nielsen2006}. 

{\it HP} and {\it HPR}  have been previously associated with lipids \cite{Guthrie2012}.
  
 \item  {\it PPARA} gene encodes the transcription factor peroxisome proliferator-activated receptor alpha (PPAR-alpha), a major regulator of lipid metabolism in the liver. PPAR-alpha serves as cellular receptor for fibrates, an anti-dyslipidemia drug that effectively lower serum triglycerides and raise serum HDL-cholesterol levels \cite{Staels2008}.

\end{itemize}

\bibliography{draft1_ref}

